\documentclass[10pt,twocolumn,twoside]{IEEEtran}
\usepackage{amsbsy,amsmath,amsfonts,amssymb,amsbsy,subfigure,booktabs,multirow}
\usepackage{bm,cite,graphicx,psfrag,pstricks,theorem,tikz,times,url,verbatim}
\usetikzlibrary{shapes,snakes,calendar,matrix,backgrounds,folding}
\usepackage[english]{babel}
\usetikzlibrary{arrows,automata}
\usetikzlibrary{calc} 
\usetikzlibrary{decorations.pathmorphing} 

\newcommand{\bi}{\begin{itemize}}
\newcommand{\ei}{\end{itemize}}
\newcommand{\ben}{\begin{enumerate}}
\newcommand{\een}{\end{enumerate}}

\newcommand{\bc}{\begin{cases}}
\newcommand{\ec}{\end{cases}}
\newcommand{\bd}{\begin{description}}
\newcommand{\ed}{\end{description}}

\newcommand{\be}{\begin{equation}}
\newcommand{\ee}{\end{equation}}
\newcommand{\bea}{\begin{eqnarray}}
\newcommand{\eea}{\end{eqnarray}}

\newcommand\T{\rule{0pt}{2.0ex}}
\newcommand\B{\rule[-0.8ex]{0pt}{0pt}}

\newtheorem{thm}{Theorem}

\newtheorem{propos}{Proposition}
\newtheorem{corol}{Corollary}

\theoremstyle{plain}
\newtheorem{remark}{Remark}

\newcommand{\nn}{\nonumber}

\graphicspath{%
     {./Figures/}
}

\begin{document}

\title{
Cross-layer design of distributed sensing-estimation  with quality feedback, Part I: Optimal schemes
}
\author{Nicol\`{o}~Michelusi~and~Urbashi~Mitra
\thanks{Copyright (c) 2014 IEEE. Personal use of this material is permitted. However, permission to use this material for any other purposes must be obtained from the IEEE by sending a request to pubs-permissions@ieee.org.}
\thanks{N. Michelusi and U. Mitra are with the Department of Electrical Engineering, University of Southern California. email addresses: \{michelus,ubli\}@usc.edu.}
\thanks{
This research has been funded in part by the following grants:
ONR N00014-09-1-0700, CCF-0917343, CCF-1117896, CNS-1213128, AFOSR FA9550-12-1-0215, and DOT CA-26-7084-00.
N. Michelusi is in part supported by AEIT (Italian association of electrical engineering) through the scholarship "Isabella Sassi Bonadonna 2013".
}
\thanks{Parts of this work have appeared in \cite{MicheAllerton,MicheGlobalsip}.}
\vspace{-5mm}}
\maketitle
\noindent\begin{abstract}
This two-part paper presents a feedback-based cross-layer framework for distributed sensing and estimation of a dynamic process by a wireless sensor network (WSN).  Sensor nodes wirelessly communicate measurements to the fusion center (FC). Cross-layer factors such as packet collisions and the sensing-transmission costs are considered. Each SN adapts its sensing-transmission action based on its own local observation quality and the estimation quality feedback from the FC under cost constraints for each SN. In this first part, the optimization complexity is reduced by exploiting the \emph{statistical symmetry} and \emph{large network approximation} of the WSN. Structural properties of the optimal policy are derived for a \emph{coordinated} and a \emph{decentralized} scheme. It is proved that a dense WSN provides \emph{sensing diversity}, so that only a few SNs with the best local observation quality need to be activated, despite the fluctuations of the WSN. The optimal policy dictates that, when the estimation quality is poor, only the \emph{best} SNs activate, otherwise all SNs remain idle to preserve energy. The costs of coordination and feedback are evaluated, revealing the scalability of the decentralized scheme to large WSNs, at the cost of performance degradation. Simulation results demonstrate cost savings from 30\% to 70\% over a non-adaptive scheme, and significant gains over a previously proposed estimator which does not consider these cross-layer factors.
 \end{abstract}
\vspace{-5mm}
\section{Introduction}
 Wireless sensor networks (WSNs) enable the monitoring of large areas via many low powered sensor nodes (SNs)
with data acquisition, processing and communication capabilities \cite{Romer}.
However, WSN design is challenged by
the high optimization complexity typical of multi-agent systems~\cite{Bernstein},
necessitating decentralized SN operation based on \emph{local} information and limited feedback,
and needs to explicitly consider the resource constraints of SNs.

In this two part paper, we present a {feedback-based} cross-layer framework for distributed sensing and estimation of a time-correlated random process at a fusion center (FC),
based on noisy measurements collected from nearby SNs,
which accounts for cross-layer factors such as the shared wireless channel, resulting in collisions among SNs, the sensing and transmission costs,
and the \emph{local} state and local view of the SNs.
{In order to cope with the uncertainties and stochastic dynamics introduced by these cross-layer components,}
the FC broadcasts feedback information to the SNs, based on the estimation quality achieved, thus enabling 
adaptation of their sensing-transmission action.
We design joint sensing-transmission policies with
the goal to minimize the mean squared estimation error (MSE) at the FC,
under a constraint on the sensing-transmission cost incurred by each SN.
 The optimal policy dictates that, when the estimation quality is poor, only the SNs with the best quality activate to improve the estimation quality at the FC, otherwise all SNs remain idle to preserve energy, at the cost of estimation quality degradation.

This first part provides a theoretical foundation for the reduction of the system complexity,
arising from the local asymmetries
 due to the decentralized operation of SNs, their local state and local view, and the multi-agent nature of the system,
 whereas Part II \cite{MichelusiP2}, informed by this theory,
investigates the design of practical schemes with low complexity.
  If one had to optimize and operate  the system under these asymmetries, the complexity would be enormous,
  \emph{i.e.}, exponential in the number of SNs, since a policy would need to be defined for each SN, and jointly optimized based on the specific local statistical properties of each SN.

We achieve complexity reduction and derive structural properties of the optimal policy by exploiting the \emph{statistical symmetry}
   and the \emph{large network approximation}. \emph{Statistical symmetry} consists in the fact that,
  despite the fluctuations in the local state of the SNs and the resulting asymmetries across the WSN,
all SNs locally experience, in the long-term, the same statistical view of the system.   
  The design implication is \emph{policy symmetry}, \emph{i.e.}, all SNs can employ a common policy to map
 their local state to a sensing-transmission action, thus significantly reducing the policy space and the optimization complexity.
 An example of statistical symmetry arises in a target tracking application: SNs closer to the target can estimate its position more accurately, whereas SNs farther away estimate it with poor accuracy; statistical symmetry implies that,
  as the target moves around within the sensing area along its trajectory, and as we consider a large number of instances of these trajectories
  in different time frames, the subset of SNs close to the target varies over time but, in the long-term, 
assuming "good" placement of the SNs (a survey on this topic is presented in \cite{Younis}),
   the statistic of the distance to the target experienced by each SN is the same for all SNs.
   
 On the other hand, the \emph{large network} approximation implies that
a large number of SNs are deployed, so that a sufficiently large (with respect to the channel/energy resource constraints of the system) set of SNs
can sense the underlying process with high accuracy in each slot, despite the temporal and spatial fluctuations in the \emph{local} accuracy state experienced across the WSN. Equivalently, in the target tracking application, there is a sufficiently large pool of SNs close to the target, which can thus estimate its position accurately.
The design implication is \emph{sensing diversity}, \emph{i.e.}, 
due to the constraints resulting from cross-layer factors such as the limited channel shared among SNs and the finite transmission resources available to the SNs,
only a few SNs with the best accuracy state need to be activated, so that the local accuracy fluctuations across the WSN can be neglected, with a consequent reduction of the state space and of the optimization complexity.
We analytically and numerically show that this approximation performs well in small-medium sized WSNs as well.

Despite the complexity reduction, the DP algorithms developed in Part I still have high complexity.
Therefore,  the aim of Part II is to design \emph{myopic policies} based on the structural properties derived in Part I,
which can be implemented with lower complexity and achieve near-optimal performance
 (no performance degradation with respect to the DP policies has been observed in our numerical evaluations).
We consider a \emph{coordinated scheme}
where the FC centrally activates each SN, and 
a \emph{decentralized scheme}, where the SNs activate in a decentralized fashion,
based on the feedback information and on their local accuracy state.
Our analysis and numerical comparison against a technique proposed in~\cite{Msechu},
which does not include these cross-layer factors,
  reveal the importance of a \emph{cross-layer approach} in the design of WSNs,  
 and of \emph{adaptation enabled by FC feedback} to cope with the consequent uncertainties and stochastic dynamics.

The problem of decentralized estimation and detection has seen a vast research effort in the last decade,
especially in the design of optimal schemes 
for parameter estimation \cite{Xiao,Thatte,Xiao2}, hypothesis testing \cite{Ray,Tsitsiklis,Chamberland}, tracking \cite{Saber,Epstein} and random field estimation \cite{Fang}.
{Distributed estimation in bandwidth-energy constrained environments has been considered in \cite{Chieh,Ribeiro,Msechu,Junlin}, for a static setting.}
 Estimation and detection problems exploiting
feedback information from the FC have been investigated in \cite{Dogandzic,Peng,Kreidl,Dey},
\emph{e.g.}, enabling adaptation of the SNs' quantizers in the estimation of a finite state Markov chain \cite{Dey}.
A consensus based approach for distributed multi-hypothesis testing has been studied in~\cite{Saligrama}.

Differently from these works, we employ a cross-layer perspective, \emph{i.e.}, we jointly consider and optimize the resource constraints typical of WSNs,
such as the shared wireless channel, resulting in collisions among SNs, the time-varying sensing capability of the SNs, their decentralized decisions,
and the cost of sensing and data transmission, and {propose a feedback mechanism from the FC to enable} \emph{adaptation} and cope with the random fluctuations  in the overall measurement quality collected at the FC,
induced by these cross-layer factors.
This is in contrast to, \emph{e.g.}, \cite{Dey}, where adaptation serves to cope with the distortion introduced by quantization.
We do not consider the problem of quantizer design,
and focus instead on a  \emph{censoring} approach \cite{Appadwedula,Msechu},
 \emph{i.e.}, quantization is fixed and sufficiently fine-grained, so that the measurements received at the FC can be approximated as Gaussian.
In fact, in light of our cross-layer design perspective, quantization may be less relevant due to the overhead required to perform essential tasks such as
synchronization and channel estimation~\cite{Appadwedula}. 

Distributed Kalman filtering for WSNs has been proposed in \cite{Olfati},
using a consensus approach and local Kalman filters at each SN. 
In this paper, Kalman filtering is employed only at the FC, which collects unfiltered observations from the SNs.
In fact, due to the poor estimation capability of SNs and their energy constraints, which force them to 
 remain idle most of the time,
the performance gain achievable by exploiting the time-correlation via local Kalman filtering may be small.

\begin{figure}
    \centering
\scalebox{0.5}{
\begin{tikzpicture}
\draw [ultra thick, ->] (0,0) -- (-0.7,-0.35);
\node at (-1.6,-.5) {Estimate $\hat X_{k}$};
\draw [ultra thick, ->] (3,0) -- (0+0.5,0);
\draw [ultra thick, ->] (-2.12,2.12) -- (0-0.3535,0+0.3535);
\draw [ultra thick, ->] (0,-3) -- (0,0-0.5);
\node at (6,-2) {Process $X_k$};
\draw [fill=black,draw=black,thick,text=white] (0,0) circle [radius=0.5];
\draw [fill=gray,draw=black,thick,text=white] (3,0) circle [radius=0.5];
\node at (3,0) {SN1};
\node at (4,0) {$Y_{1,k}$};
\draw [ultra thick, ->] (3.7,0) -- (3.5,0);
\def\x{3.5}
\def\y{0.35}
\draw [fill=white, ultra thick] (2.12,2.12) circle [radius=0.5];
\node at (2.12,2.12) {SN2};
\node at (2.83+0.2,2.83-0.2) {$Y_{2,k}$};
\draw [ultra thick,dashed, ->] (0,0) -- (2.47-0.7,2.47-0.7);
\draw [ultra thick, ->] (2.62,2.62) -- (2.47,2.47);
\def\x{2.12+0.6}
\def\y{2.12-0.25}
\draw [fill=white, ultra thick] (0,3) circle [radius=0.5];
\node at (0,3) {SN3};
\node at (0,4) {$Y_{3,k}$};
\draw [ultra thick,dashed, ->] (0,0) -- (0,2.5);
\draw [ultra thick, ->] (0,3.7) -- (0,3.5);
\def\x{0.6}
\def\y{3-0.25}
\draw [fill=gray,draw=black,thick,text=white] (-2.12,2.12) circle [radius=0.5];
\node at (-2.12,2.12) {SN4};
\node at (-2.83,2.83) {$Y_{4,k}$};
\draw [ultra thick, ->] (-2.62,2.62) -- (-2.47,2.47);
\def\x{-2.12-1.2-0.6}
\def\y{2.12-0.25}
\draw [fill=white, ultra thick] (-3,0) circle [radius=0.5];
\node at (-3,0) {SN5};
\node[above] at (-1.5,0) {$\mathbf D_{k+1}$};
\node at (-4,0) {$Y_{5,k}$};
\draw [ultra thick,dashed, ->] (0,0) -- (-2.5,0);
\draw [ultra thick, ->] (-3.7,0) -- (-3.5,0);
\def\x{-3.5-1.2}
\def\y{0.35}
\draw [fill=white, ultra thick] (-2.12,-2.12) circle [radius=0.5];
\node at (-2.12,-2.12) {SN6};
\node at (-2.83-0.2,-2.83+0.2) {$Y_{6,k}$};
\draw [ultra thick,dashed, ->] (0,0) -- (-2.47+0.7,-2.47+0.7);
\draw [ultra thick, ->] (-2.62,-2.62) -- (-2.47,-2.47);
\def\x{-2.12-1.2-0.6}
\def\y{-2.12-0.25}
\draw [fill=gray,draw=black,thick,text=white] (0,-3) circle [radius=0.5];
\node at (0,-3) {SN7};
\node at (0,-4) {$Y_{7,k}$};
\draw [ultra thick, ->] (0,-3.7) -- (0,-3.5);
\def\x{0-1.2-0.6}
\def\y{-3-0.25}
\draw [fill=white, ultra thick] (2.12,-2.12) circle [radius=0.5];
\node at (2.12,-2.12) {SN8};
\node at (2.83,-2.83) {$Y_{8,k}$};
\draw [ultra thick,dashed, ->] (0,0) -- (2.47-0.7,-2.47+0.7);
\draw [ultra thick, ->] (2.62,-2.62) -- (2.47,-2.47);
\def\x{2.12+0.6}
\def\y{-2.12-0.25}
\node [text=white] at (0,0) {FC};
\draw[->,decorate, decoration={snake, segment length=7mm, amplitude=1mm}] (5,-1.7) -- (4,-0.3);
\draw[->,decorate, decoration={snake, segment length=7mm, amplitude=1mm}] (4.4,-2) -- (3,-2.5);
\draw[->,decorate, decoration={snake, segment length=7mm, amplitude=1mm}] (6,-1.7) -- (3.6,2.4);
\end{tikzpicture}}
\vspace{-3mm}
\caption{A WSN for distributed estimation, with FC quality feedback.
Each SN decides to either remain idle with cost $0$ or to collect and transmit to the FC the measurement $Y_{n,k}$ of $X_k$ with local measurement SNR $S_{M,n,k}$ and cost $c_{\mathrm{TX}}+\phi S_{M,n,k}$. The shared wireless channel results in collisions and packet losses. The FC, 
based on the measurements received, computes an MMSE estimate of $X_k$, $\hat X_k$, and broadcasts the instruction $\mathbf D_{k+1}$ based on the estimation quality 
achieved,
which is used by the SNs to adjust their sensing-transmission parameters for the next slot.}
\label{fig:WSN}
\vspace{-5mm}
\end{figure}
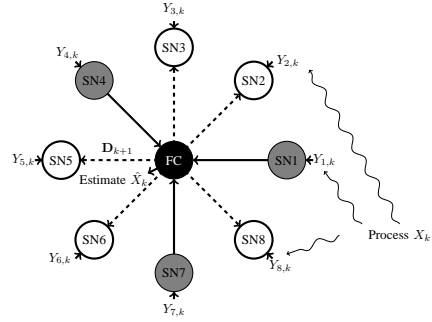

 This paper is organized as follows.
  In Sec.~\ref{probform}, we motivate our approach and summarize the main results.
  In Sec.~\ref{sysmo}, we present the system model and the optimization problem.
 In Sec.~\ref{analysis}, we present
 the analysis of the coordinated and decentralized schemes.
In Sec.~\ref{numres}, we provide numerical results.
In Sec.~\ref{conclusions}, we conclude the paper.
The analytical proofs are provided in the Appendix.
\vspace{-0.3cm}
\section{Motivation}
\label{probform}
 Consider a WSN, depicted in Fig.~\ref{fig:WSN}, with one FC, whose goal
 is to track a \emph{stationary Markov} process $\{X_k,\ k\geq 0\}$,
 based on measurements collected  by $N_S$ nearby SNs.
 The probability density function (if $X_k$ is continuous, or probability mass function, if $X_k$ is discrete) of $X_{k+1}$ given $X_k$ is denoted as 
 $p_{X}(X_{k+1}|X_k)$. In this paper, we consider the  scalar linear Gaussian state space model
 \begin{align}
\label{markovstate}
X_{k+1}=\sqrt{\alpha}X_k+Z_k,
\end{align}
where $k\in\mathbb N\equiv\{0,1,2,\dots\}$ is the slot index,
 $\alpha\in [0,1)$ is the \emph{time-correlation parameter}  and $Z_k\sim\mathcal N(0,\sigma_Z^2)$,
 so that $X_{k+1}|X_k\sim\mathcal N(\sqrt{\alpha} X_k,\sigma_Z^2)$.
 This model arises, for instance, in temperature tracking applications, where $X_k$ represents the temperature fluctuations around its mean \cite{Tandeo}.
 We denote the statistical power of $X_k$ as $\sigma_X^2=\frac{\sigma_Z^2}{1-\alpha}$,
 and assume  $\sigma_X^2=1$, since any other value can be obtained by scaling.
 
Each SN incurs the transmission cost $c_{\mathrm{TX}}$ to report its measurement to the FC.
   The $N_S$ SNs share 
 a set of $B{\leq}N_S$ orthogonal single-hop wireless channels to report their measurements to the FC.
 We employ the collision channel model, \emph{i.e.},
 the transmission on a given channel is successful if and only if one SN transmits in that channel. 
 This model is commonly employed in the analysis of multi-access communication schemes,
and lends itself to analysis.\footnote{Other channel models can be accommodated by defining, more generally, a probability mass function (PMF) $
 p_{R|T}(r|t)\triangleq\mathbb P(R_k=r|T_k=t),\ r\in\{0,1,\dots,t\}$, where
 $T_k$ and $R_k$ are the number of SNs that transmit and of packets successfully received at the FC, respectively.}
 
Referring to the model (\ref{markovstate}),
 assume for simplicity $B{=}1$
 and that each SN measures $X_k$ noiselessly (the noisy case with $B{\geq}1$ is considered in the rest of the paper).
   Let $O_{n,k}$ be the \emph{transmission outcome} for SN $n$, \emph{i.e.}, $O_{n,k}{=}1$ if and only if its transmission is successful. 
    Then, if at least one measurement is collected at the FC, \emph{i.e.}, $\prod_n(1{-}O_{n,k}){=}0$,
 the MSE is $0$. On the other hand, if no measurements are successfully received, \emph{i.e.}, $\prod_n(1{-}O_{n,k}){=}1$, then
$X_k$ is estimated via prediction. Therefore, if the transmission has been successful in slot $k{-}J_k{-}1$, for some $J_k{\geq}0$,
so that $X_{k-J_k-1}$ is perfectly known at the FC,
 but transmission failures or no transmission attempts occurred in slots $k{-}J_k,k{-}J_k{+}1,\dots,k$, then $X_k$ is estimated as $\hat X_k=\sqrt{\alpha}^{J_k+1}X_{k-J_k-1}$ and
 the MSE at the end of slot $k$ is $(1{-}\alpha^{J_k+1})$.
 Due to the decentralized sensing-transmission decision of the SNs and the
  shared wireless channel, which may result in collisions among SNs, random and unpredictable fluctuations in the transmission outcome $O_{n,k}$ may occur at the FC,
  so that the MSE evolves randomly over time. In order to control the uncertainty and system dynamics introduced by these cross-layer factors, we thus propose a
  feedback-based adaptive  scheme where
 the SNs adapt their activation strategy over time, \emph{i.e.}, whether to sense-transmit their measurement with cost $c_{\mathrm{TX}}$ (denoted as $A_{n,k}=1$) or remain idle 
 with no cost (denoted as $A_{n,k}=0$), based on \emph{quality feedback} from the FC, captured by the state variable $J_k$.
 The goal is to design the activation policy so as to minimize the expected MSE
$\bar M\triangleq\mathbb E[\prod_n(1-O_{n})(1-\alpha^{J+1})]$
 at the FC,\footnote{The slot index $k$ is removed for simplicity to denote steady-state regime.}
under SN sensing-transmission cost constraints, $\bar C_n=\mathbb E[A_{n}c_{\mathrm{TX}}]\leq \epsilon/N_S,\ \forall  n$.
We consider the following schemes.
\vspace{-3mm}
\subsection{Coordinated scheme}
\label{coordscheme}
 In this scheme, the FC centrally schedules the activation $A_{n,k}$ of  each SN.
One design approach to optimize the MSE
is to maximize the number of measurements collected at the FC in each slot, under the cost constraint for each SN.
This is denoted as \emph{max aggregate SNR scheme} (MAX-SNR) in the rest of the paper.
If $\epsilon\geq c_{\mathrm{TX}}$, the optimal strategy dictates to activate randomly one and only one SN in each slot,
resulting in a successful transmission, hence the MSE is $0$ in each slot (Theorem \ref{thm1}). We thus have
 $\bar M=0$, $\bar C_n{=}\frac{c_{\mathrm{TX}}}{N_S}{\leq}\frac{\epsilon}{N_S}$, hence
 a \emph{non-adaptive} scheme is optimal in this case.
 \vspace{-3mm}
 \subsection{Decentralized scheme}
 \label{distscheme}
\noindent  Unfortunately, the coordinated scheme is not scalable to large WSNs,
 due to the centralized scheduling performed by the FC.
 Therefore, a decentralized approach, where the SNs make local decisions, leveraging only local information
 and minimal feedback information, is more practical.
 We thus devise
a decentralized scheme, where each SN activates with common probability $q_k$ in slot $k$.
 Following the same design principle of optimizing the expected number of measurements collected at the FC (MAX-SNR scheme),
we define a  \emph{non-adaptive} (NA) scheme where each
 SN activates with probability $q_k{=}\zeta/N_S$ in each slot, where we have defined the normalized transmission probability per channel $\zeta{\in}[0,N_S]$.
In this case, $\{J_k\}$ is a Markov chain.
Using the \emph{large network approximation} $N_S{\gg}1$ with fixed $\zeta$,
its transition probabilities are
$\mathbb P(J_{k+1}{=}j{+}1|J_k{=}j){\simeq}1{-}\zeta e^{-\zeta}$,
 $\mathbb P(J_{k+1}{=}0|J_k{=}j){\simeq}\zeta e^{-\zeta}$,
 and the steady-state probability of $J_k{=}j$ is given by   $\pi_J(j){\simeq}\zeta e^{-\zeta}(1-\zeta e^{-\zeta})^j,j{\geq}0$.
 By averaging over $\pi_J(j)$,
 the average SN cost and MSE
 are
 \begin{align}
\bar C_n^{(NA)}=\frac{\zeta}{N_S}c_{\mathrm{TX}},\ 
\bar M^{(NA)}=\frac{(1-\alpha)(1-\zeta e^{-\zeta})}{1-\alpha+\alpha\zeta e^{-\zeta}}.
\label{NA}
 \end{align}
 
 \begin{table*}[t]

\caption{Main system parameters}
\vspace{-5mm}
\label{tab1}
\begin{center}
\footnotesize
\scalebox{0.88}{
\begin{tabular}{|c| l | c | l | c | l | c | l |}
\hline\T\B $\{X_k\}$& random process to be tracked
&
$S_A$& local ambient SNR
 &
$Y_{n,k}$& measurement of SN $n$ in slot $k$
&
$\gamma_{n,k}$& accuracy state with 
 s.s.d. $\pi_{\gamma}(\gamma)$
  \\\hline\T\B 
  $\alpha$& time-correlation parameter
&$S_{M,n,k}$& local measurement SNR
&$A_{n,k}$& activation of SN $n$, slot $k$
&$B_{n,k}$& channel ID for SN $n$, slot $k$
\\\hline\T\B
 $\Lambda_k$& aggregate SNR at FC
& $\phi S_{M,n,k}$& sensing cost
& $c_{\mathrm{TX}}$& transmission cost
 &$B$& \# channels available, $B\leq N_S$
 \\\hline\T\B
$V_k$ & prior variance 
&$\hat V_k$ & posterior variance 
&$q$ & SN activation  probability 
& $N_S$ & \# of SNs, $N_S\geq B$
\\\hline\T\B 
$\theta{\triangleq}\frac{\phi}{c_{\mathrm{TX}}}$& normalized unitary sensing cost &
$\bar M_{\delta}$ & average MSE
&
$\bar C_{\delta}^{n}$ & 
\multicolumn{3}{l|}{
average sensing-transmission cost of SN $n$
}
\\
\hline
\end{tabular}
}
\end{center}
\vspace{-5mm}
\end{table*}

\begin{figure}[t]
\centering
\includegraphics[width = .8\linewidth,trim = 10mm 4mm 10mm 9mm,clip=true]{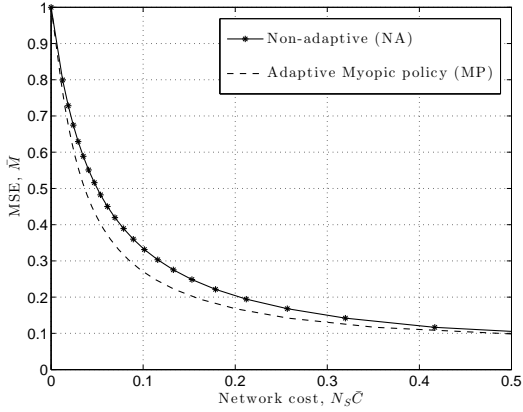}
\vspace{-3mm}
\caption{Trade-off between network cost and MSE. $\alpha=0.95$, $c_{\mathrm{TX}}=1$.}
\label{TOYEX}
\vspace{-5mm}
\end{figure}

 Unfortunately, this design approach
 fails to achieve good performance in general,
since the decentralized SN activation and the collisions among SNs result in
random fluctuations in the number of measurements collected at the FC (which may be zero in case of collisions),
hence high uncertainty and poor MSE performance.
In order to control the uncertainty in the system, we propose an adaptive scheme
   where the activation probability $q_k$ is adapted over time by the FC, based on the current quality
  state $J_k$. Such adaptive policy is denoted as $q(\cdot){=}\zeta(\cdot)/N_S$.
    The value of the activation probability $q_k{=}q(J_k){=}\zeta(J_k)/N_S$ is broadcasted by the FC at the beginning of each slot.
    In particular, consider the myopic policy (MP), which 
   determines $q_k{=}q_{MP}(j){=}\zeta_{MP}(j)/N_S$ in state $J_k{=}j$ so as to 
    optimize a trade-off between the instantaneous 
    expected MSE and the cost for each SN,
    \begin{align}
    \label{MP}
 \zeta_{MP}(j)=\arg\min_\zeta (1-\zeta e^{-\zeta})(1-\alpha^{j+1})+\lambda\zeta,
    \end{align}
    where $\lambda{\geq}0$ captures the desired trade-off, $(1{-}\zeta e^{-\zeta})$ is the probability that 
    no measurements are received at the FC, and $(1{-}\alpha^{j+1})$ is the corresponding MSE achieved.
    The solution to this optimization problem is studied in Part II, and is given by
    $\zeta_{MP}(j){=}0$ if  $\lambda{\geq}(1{-}\alpha^{j+1})$,
    otherwise it is the unique $\zeta\in[0,1]$ solution of
    $e^{-\zeta}(1{-}\alpha^{j+1})(1{-}\zeta){=}\lambda$ (see  \cite[Corollary~2]{MichelusiP2}).    
    Using the bound $e^{-\zeta}{\leq}1$ in (\ref{MP}),
    we obtain the approximate MP (AMP), upper bound to $\zeta_{MP}(j)$,
    \begin{align}
    \label{AMP}
  \zeta_{AMP}(j)=\left[1-\frac{\lambda}{1-\alpha^{j+1}}\right]^+\geq \zeta_{MP}(j),\ \forall j.
    \end{align}
The AMP
    $\zeta_{AMP}(j)$ is an increasing function of $j$, 
    \emph{i.e.}, the higher the uncertainty (the larger $J_k{=}j$), the higher the activation probability,
    which approaches $\left[1-\lambda\right]^+$ for $j\to\infty$.
\emph{Hence, AMP has the desirable property that,
the higher the uncertainty in the current estimate, the more the SNs are 
incentivized to activate, at higher cost, in order to estimate $X_k$ accurately at the FC.}
Under AMP, we have
    \begin{align}
        \pi_J(j)=\frac{\prod_{i=0}^{j-1}(1-\zeta_{AMP}(i) e^{-\zeta_{AMP}(i)})}{
        \sum_{l=0}^{\infty}\prod_{i=0}^{l-1}(1-\zeta_{AMP}(i) e^{-\zeta_{AMP}(i)})
        },\ j\geq 0,
    \end{align}
so that  the average SN cost and MSE
 are given by
 \begin{align*}
 \left\{
\begin{array}{ll}
\!\!\!\!\bar C_n^{(AMP)}&\!\!\!\!\!\!=\!
\sum_{j=0}^{\infty}\pi_J(j)\frac{\zeta_{AMP}(j)}{N_S}c_{\mathrm{TX}},\\
\!\!\!\!\bar M^{(AMP)}&\!\!\!\!\!\!=\!
\sum_{j=0}^{\infty}\pi_J(j)(1-\zeta_{AMP}(j) e^{-\zeta_{AMP}(j)})(1\!-\!\alpha^{j+1}).\!
\end{array}\right.
 \end{align*}
 By varying $\zeta{\in}[0,1]$ in (\ref{NA}), $\lambda{\geq}0$ in (\ref{AMP}), we obtain
  the cost-MSE  trade-off depicted in Fig.~\ref{TOYEX},
 which shows that AMP reduces the sensing-transmission cost for each SN by 30\% with respect to NA.
 Therefore,
  adaptation
 to the quality state yields performance gains in the cost-MSE trade-off and 
 effectively copes with the uncertainty introduced by the network and cross-layer components.
 
In the next sections, we will extend the analysis to the more general case $B{\geq}1$,
where the SNs collect noisy measurements of $X_k$, whose quality is affected by an internal \emph{accuracy state} evolving as a Markov chain,
and by control performed by each SN.
In Theorem~\ref{thm1},
we will show that,
in the coordinated scheme,
 the MAX-SNR scheme, which maximizes the 
expected aggregate SNR collected at the FC in each slot under the SN cost constraint,
is optimal in the \emph{best}-$\gamma$  scenario, where
all SNs have deterministically the best accuracy state,
under some conditions on the maximum cost for each SN, $\epsilon/N_S$.
In Theorem~\ref{thm2}, we will show that this strategy is near-optimal for \emph{large networks} in the \emph{Markov}-$\gamma$ scenario,
where the accuracy state of each SN follows a Markov chain.
For the decentralized scheme, we derive structural properties
and exploit the large network approximation to design a DP algorithm with lower complexity. 
In Part II, we will further investigate the design of myopic policies for this more general setting,
for both the coordinated and decentralized schemes.

 \begin{remark}
This framework and the following analysis can also be applied  to other time-correlated signals, \emph{e.g.}, the two state Markov chain
 \begin{align}
 X_{k+1}=X_k\oplus Z_k,\ X_{k},Z_k\in\{0,1\},
 \end{align}
 where $\oplus$ denotes the sum modulo 2,
  and $Z_k$ has distribution $p_{Z|X}(z|x){\triangleq}\mathbb P(Z_k{=}z|X_k{=}x),z,x{\in}\{0,1\}$.
 This model  arises, for instance, in spectrum sensing applications \cite{MicheISIT},
 where $X_k$ denotes the channel occupancy state ($X_k{=}0$ if idle, $X_k{=}1$ if busy).
In this case, the quality feedback is captured by the log-likelihood ratio $\ln(\mathbb P(X_k{=}1|\text{history})|\mathbb P(X_k{=}0|\text{history}))$,
 reflecting the current detection accuracy, and the expected MSE can be replaced with the expected detection error probability.
 
The model (\ref{markovstate}) can also be extended to the multi-dimensional case, \emph{e.g.},
 in target tracking applications where the vector $X_k$ represents the position and speed in slot $k$.
 In this case, $\sqrt{\alpha}$ is replaced by a proper matrix \cite{Xiao}, and the feedback quality is represented by the error covariance matrix (or by its trace, for
 dimensionality reduction purposes).
 \end{remark}
\section{System Model and Optimization Problem}
\label{sysmo}
 In this section, we present the system model. The main parameters are listed in Table \ref{tab1}.
Time is slotted and all SNs are assumed to be perfectly synchronized.\footnote{Note, however, that 
we also presume random access, which allows for some robustness against imperfect synchronization \cite{Gaudenzi}.}
 Each slot includes three~phases:
 \begin{enumerate}
 \item \emph{FC instruction} $\mathbf D_k$, broadcasted by the FC (Sec.~\ref{FCinstruct});
 \item \emph{Sensing and transmission to FC}: given $\mathbf D_k$, each SN selects its
 sensing-transmission action (Sec. \ref{ph2});
 \item \emph{Estimation at FC}: given the measurements collected, the FC estimates $X_k$ via Kalman filtering (Sec.~\ref{P3}).
 \end{enumerate}
 \vspace{-5mm}
\subsection{Sensing and transmission to FC}
 \label{ph2}
\noindent  Each SN, at the beginning of slot $k$, given the instruction $\mathbf D_k$ broadcasted by the FC,
  selects (possibly, in a randomized fashion) 
  the sensing-transmission parameters $(A_{n,k},S_{M,n,k},B_{n,k})$, 
 where $A_{n,k}{\in}\{0,1\}$ is the \emph{activation} decision of SN $n$,
 $S_{M,n,k}{\geq}0$ is the \emph{local measurement SNR} specified below, and $B_{n,k}{\in}\{0,1,2,\dots,B\}$ is the \emph{channel index}.
 If $A_{n,k}{=}0$, SN $n$ remains idle, hence $S_{M,n,k}{=}0$ (no measurement collected) and $B_{n,k}{=}0$ (no channel selected).
On the other hand, if $A_{n,k}{=}1$, then $B_{n,k}{\in}\{1,2,\dots,B\}$ and
 the measurement of $X_k$ collected by SN $n$ is given by
\begin{align}
\label{Ynk}
Y_{n,k}=\gamma_{n,k}X_{k}+W_{A,n,k}+W_{M,n,k},
\end{align}
where $W_{A,n,k}{\sim}\mathcal N(0,1/S_A)$ is the \emph{ambient noise},
and $W_{M,n,k}{\sim}\mathcal N(0,1/S_{M,n,k})$ is the \emph{measurement noise} introduced by the sensing apparatus,
independent of each other, over time and across SNs,
  $S_{A}$ is the  \emph{local ambient SNR}, and $S_{M,n,k}$ is the
   \emph{local measurement SNR}, controlled by the $n$th SN, resulting in the sensing cost $\phi S_{M,n,k}$, for some $\phi\geq 0$.
 Note that this assumption is practical. For instance,
 SN $n$ may compute an average from a controlled number $M_{n,k}$ of independent measurements, each with
fixed ambient noise and i.i.d. measurement noise with  variance $\sigma_M^2$
 and cost $c_S$,
 resulting in the local measurement SNR $S_{M,n,k}=M_{n,k}/\sigma_M^2$ and in the overall
 sensing cost $c_SM_{n,k}=(c_S\sigma_M^2)S_{M,n,k}$.

We assume that
   a fixed quantization scheme is employed, \emph{i.e.}, a fixed number of bits is transmitted to the FC,\footnote{Therefore, 
   the ambient SNR $S_A$ and noise $W_{A,n,k}$ can also be interpreted, respectively, as the quantization SNR floor and the Gaussian approximation of the quantization error.}
and that each SN is unaware of its own distance to the FC
and it does not employ power adaptation, but it
 transmits with constant power, so as to provide a given coverage requirement,
resulting in the overall transmission cost $c_{\mathrm{TX}}$, common to all SNs.
 The FC is assumed to be within the coverage area 
of each SN. A varying $c_{\mathrm{TX}}$ can be easily incorporated with increased book-keeping.
We define the \emph{normalized unitary sensing cost} $\theta\triangleq\frac{\phi}{c_{\mathrm{TX}}}$.
No cost is incurred if the SN remains idle.
The overall sensing-transmission cost
is thus $c_{SN}(A_{n,k},S_{M,n,k}){\triangleq}A_{n,k}(c_{\mathrm{TX}}+\phi S_{M,n,k})$.
We define the \emph{sample average sensing-transmission cost for SN $n$} over a time horizon of length $T+1$ as
\begin{align}
\label{ctn}
C_n^{T}(A_{n,0}^T,S_{M,n,0}^T)=\frac{1}{T+1}\sum_{k=0}^{T}c_{SN}(A_{n,k},S_{M,n,k}).
\end{align}

The \emph{accuracy state} $\gamma_{n,k}$, taking values in the finite set $\Gamma$, models the ability of SN $n$ to accurately measure $X_k$.
We model it as a Markov chain with transition probability $\mathbb P(\gamma_{n,k+1}{=}\gamma_2|\gamma_{n,k}{=}\gamma_1){=}P_\gamma(\gamma_1;\gamma_2)$ 
and steady-state distribution $\pi_{\gamma}(\gamma)$,
i.i.d. across SNs, and we let $\boldsymbol{\gamma}_k=(\gamma_{1,k},\gamma_{2,k},\dots,\gamma_{N_S,k})$.
{Such a model arises, \emph{e.g.}, in a target tracking application, where the power of the received signal diminishes with the distance, which 
evolves following Markov dynamics as a function of the relative motion of the SN and the target  \cite{Xiao}.
The Markov assumption on $\gamma_{n,k}$ is used for analytical tractability, but the following analysis requires only the existence of
the steady-state distribution $\pi_{\gamma}(\gamma)$, and therefore it applies to non-Markov dynamics as well.}
In practice, $\gamma_{n,k}$ varies slowly over time, \emph{e.g.}, as a function of the SN position with respect to the
source of the process $X_k$, and therefore it can be tracked accurately
 from the sample mean and sample variance estimates of the measurement noise. 
  We denote the best accuracy state as $\gamma_{\max}{=}\max\{\Gamma\}$,
and, without loss of generality, we assume $\gamma_{\max}{=}1$ and $\pi_{\gamma}(\gamma_{\max}){>}0$.
 We denote the general scenario where $\gamma_{n,k}$ follows a Markov chain as \emph{Markov-}$\gamma$ scenario,
and the special cases where $\gamma_{n,k}{=}\gamma_{\max},\forall n,k$,  deterministically 
and $\gamma_{n,k}$ is i.i.d. over time as \emph{best-}$\gamma$ and \emph{i.i.d.-}$\gamma$ scenarios, respectively.
\vspace{-3mm}
\begin{remark}
\label{rem22}
 Note that the local accuracy state may vary significantly over both time and space, yielding instantaneous \emph{asymmetries} in the WSN. Typically, design of 
 asymmetric systems suffers from the curse of dimensionality. Herein, we assume that \emph{statistical symmetry} holds, in the sense that, in the long term, the SNs have the same statistical view of the system, despite the local temporal and spatial fluctuations of the state. 
 As a consequence, we assume \emph{policy symmetry}, \emph{i.e.}, all SNs employ the same policy to map
 their local state to a sensing-transmission action, thus significantly reducing the policy space and the optimization complexity.
 \end{remark}
   
\subsection{MMSE estimator at the FC via Kalman filtering}
\label{P3}
\noindent The weighted average measurement
   \begin{align}
   \label{bari}
\bar Y_{k}\triangleq\frac{\sum_{n}O_{n,k}\frac{S_{n,k}}{\gamma_{n,k}}Y_{n,k}}{\sum_{n}O_{n,k}S_{n,k}}
   \end{align}
   is a sufficient statistic for $X_k$,
   where $S_{n,k}$ is the \emph{local SNR} for SN $n$, defined as
      \begin{align}
      \label{Slocal}
   S_{n,k}=\frac{\mathbb E[(\gamma_{n,k}X_{k})^2|\gamma_{n,k}]}{\mathbb E[(W_{A,n,k}+W_{M,n,k})^2]}=
  \gamma_{n,k}^2 \frac{S_{A}S_{M,n,k}}{S_{A}+S_{M,n,k}}.
   \end{align}   
    Given the transmission outcome and $X_k$,
   $\bar Y_k$ is a Gaussian random variable with mean $X_k$ and variance
   $\Lambda_k^{-1}$,
where we have defined the \emph{aggregate SNR} collected at the FC as
\begin{align}
\label{Stot}
\Lambda_k\triangleq\sum_{n=1}^{N_S}O_{n,k}S_{n,k}.
\end{align}

 Let $\hat X_{k-1}$ and $\hat V_{k-1}$ be the posterior mean (\emph{i.e.}, the MMSE estimate)
and variance of $X_{k-1}$ at the FC at the end of slot $k-1$,
\emph{i.e.}, $X_{k-1}\sim \mathcal N(\hat X_{k-1},\hat V_{k-1})$ is the belief of $X_{k-1}$ at the FC.
 Before collecting the measurements from the SNs in slot $k$,
 using~(\ref{markovstate}),
 the belief of $X_{k}$ is $X_{k}\sim\mathcal N(\sqrt{\alpha} \hat X_{k-1},V_k)$,
 where $V_k$ is the \emph{prior variance} of $X_k$,
defined recursively as
\begin{align}
\label{nu}
&V_k=\alpha \hat V_{k-1}+\sigma_Z^2=
1-\alpha(1-\hat V_{k-1})\triangleq \nu(\hat V_{k-1}).
\end{align}
Then, upon collecting the weighted average measurement $\bar Y_k$ (\ref{bari})
 with aggregate SNR $\Lambda_k$,\footnote{We assume that each active SN, in addition to $Y_{n,k}$, also provides to the FC the value of $\gamma_{n,k}$
and $S_{n,k}$, which is employed in the Kalman filter.}
the FC updates the \emph{posterior variance} $\hat V_k$ and  mean $\hat X_{k}$ of $X_{k}$ as
\begin{align}
\label{nu2}
\left\{\begin{array}{l}
\hat V_k=\frac{V_k}{1+V_k\Lambda_k}
\triangleq \hat \nu(V_{k},\Lambda_k),
\\
\hat X_{k}=\sqrt{\alpha} \hat X_{k-1}+\Lambda_k\hat V_k\left(\bar Y_k -\sqrt{\alpha}\hat X_{k-1}\right).
\end{array}\right.
\end{align}
The function $\nu(\hat V_{k-1})$ in (\ref{nu}) determines  the prior variance of $X_k$, given the posterior variance of $X_{k-1}$,
whereas $\nu(V_k,\Lambda_k)$ in (\ref{nu2}) determines the posterior variance of $X_k$, given its prior variance $V_k$,
as a function of the aggregate SNR $\Lambda_k$ collected at the FC.
The MSE in slot $k$ is thus
\begin{align}
&\mathbb E\left[\left.(\hat X_{k}- X_{k})^2\right|V_{k},\Lambda_k\right]
=\hat \nu(V_{k},\Lambda_k).
\end{align}
We define recursively $\hat\nu^0(V_0;\Lambda_0^{0})=\hat \nu(V_{0},\Lambda_0)$
and, for $k>0$,
\begin{align}
\label{nukrec}
\hat\nu^k(V_0;\Lambda_0^{k})=\hat\nu\left(\hat\nu^{k-1}(V_0;\Lambda_0^{k-1}),\Lambda_k\right),
\end{align}
where $\Lambda_0^k=(\Lambda_0,\Lambda_1,\dots,\Lambda_k)$ is the aggregate SNR sequence collected at the FC
from slot $0$ to slot $k$. Then,
we can write $\hat V_k=\hat\nu^k(V_0;\Lambda_0^{k})$.
We define the \emph{sample average MSE} under $\Lambda_0^T$ over a time horizon of length $T+1$ as
\begin{align}
\label{RT}
R_T(V_0;\Lambda_0^T)=\frac{1}{T+1}\sum_{k=0}^{T}\hat\nu^k(V_0;\Lambda_0^{k}).
\end{align}
Note that the prior and posterior variances $V_k$ and $\hat V_k$ take value between $[0,1]$,
where the extreme values $0$ and $1$ correspond, respectively, to
 minimum ($X_k$ perfectly known) and maximum ($X_k$ is completely unknown) uncertainty.
Therefore, $R_T(V_0;\Lambda_0^T)\in [0,1],\ \forall T,V_0,\Lambda_0^T$, and the system is stable.
\vspace{-4mm}
\subsection{FC instruction policy}
\label{FCinstruct}
\begin{table}[t]
\caption{FC instruction  policy}
\vspace{-8mm}
\label{tabins}
\begin{center}
\footnotesize
\scalebox{0.83}{
\begin{tabular}{|c| c | c | c |}
\hline\T\B 
\multirow{2}{*}{\em Scheme} & \multirow{2}{*}{\em Activity $A_{n,k}$}  & \em Local measurement & \multirow{2}{*}{\em Channel ID $B_{n,k}$}\\
&  &\em SNR $S_{M,n,k}$ & 
\\\hline\T\B 
{\bf Coordinated}    & Centralized, $@$ FC & Centralized, $@$ FC & Centralized, $@$ FC
\\\hline\T\B 
\multirow{2}{*}{{\bf Decentralized}} & Local, w.p. $q_k(\omega_{n,k})$ & Local, $\sim S_{M,k}(\omega_{n,k})$ & \multirow{2}{*}{Local, random}
\\
 & $q_k(\cdot)$ given by FC & $S_{M,k}(\cdot)$  given by FC  &
\\
\hline
\end{tabular}
}
\end{center}
\vspace{-9mm}
\end{table}

\noindent At the beginning of  slot $k$,
the FC broadcasts an
\emph{instruction} $\mathbf D_k{\in}\mathcal D$, which, together with the local accuracy state $\gamma_{n,k}$,
is used by SN $n$ to select $(A_{n,k},S_{M,n,k},B_{n,k})$ as in Sec. \ref{ph2}. We consider the following schemes, summarized in Table \ref{tabins}:
\subsubsection{Coordinated scheme}
  In the coordinated scheme, given $V_k$ and $\boldsymbol{\gamma}_k$, the FC schedules the sensing-transmission action $d_{n,k}{=}(A_{n,k},S_{M,n,k},B_{n,k})$ of  each SN,  so that $\mathbf D_k{=}(d_{1,k},d_{2,k},\dots,d_{N_{S},k})$.
Note that each SN is required to report its accuracy state to the FC, whenever its value changes,
  so that $\boldsymbol{\gamma}_{k}$ is perfectly known at the FC at the beginning of slot $k$.
  Letting $\pi_{\boldsymbol{\gamma},k}$ be the belief of $\boldsymbol\gamma_{k}$ at the FC, we have that
  $\pi_{\boldsymbol{\gamma},k}(\boldsymbol\gamma){=}\chi(\boldsymbol\gamma{=}\boldsymbol\gamma_{k})$,
  where $\chi(\cdot)$ is the indicator function. 
  In Sec.~\ref{commover}, we will analyze the cost of communication overhead to keep such state information at the FC.
The value $\mathbf D_k$ is selected according to some (possibly, non-stationary) \emph{instruction policy}
 $\delta_k(\mathbf d|V_k,\pi_{\boldsymbol{\gamma},k})\triangleq\mathbb P(\mathbf D_{k}=\mathbf d|V_k,\pi_{\boldsymbol{\gamma},k})$.
 \subsubsection{Decentralized scheme}
 In the decentralized scheme, the FC specifies 
 $\mathbf D_k{=}(q_k(\cdot),S_{M,k}(\cdot))$,
  where $q_k{:}\Gamma{\mapsto}[0,1]$ and $S_{M,k}{:}\Gamma{\mapsto}[0,\infty)$
  are, respectively,
the activation probability and 
the local measurement SNR functions  employed by each SN to select their sensing-transmission strategy in a decentralized manner, as a function of the local accuracy state $\gamma_{n,k}$.
Therefore, $\mathbf D_k$ takes value in the set $\mathcal D\equiv ([0,1]^\Gamma\times\mathbb R_+^\Gamma)$,
and is generated according to some (possibly, non-stationary) policy $\delta_k(\mathbf d|V_k,\pi_{\boldsymbol\gamma,k}){\triangleq}\mathbb P(\mathbf D_{k}{=}\mathbf d|V_k,\pi_{\boldsymbol{\gamma},k})$, where
$\pi_{\boldsymbol\gamma,k}(\boldsymbol\gamma_k)=\mathbb P(\boldsymbol\gamma_k|\mathcal H_k)$
is the belief state of the accuracy state vector $\boldsymbol\gamma_k$, given the history of observations collected up to time $k$ at the FC, $\mathcal H_k$. 
  Given 
$\mathbf D_k{=}(q_k(\cdot),S_{M,k}(\cdot))$
and the local accuracy state $\gamma_{n,k}$,
 SN $n$ chooses its action $(A_{n,k},S_{M,n,k},B_{n,k})$
as $A_{n,k}{=}1$ with probability $q_k(\gamma_{n,k})$,
$A_{n,k}{=}0$ otherwise; if $A_{n,k}{=}1$, then
$S_{M,n,k}{=}S_{M,k}(\gamma_{n,k})$ and $B_{n,k}$ is chosen uniformly from the set of  channels $\{1,2,\dots, B\}$
{(if $A_{n,k}{=}0$, then $S_{M,n,k}{=}B_{n,k}{=}0$).}
Due to the randomized channel accesses, this scheme may result in collisions among SNs.
\vspace{-3mm}
\begin{remark}
\label{PMF}
The choice of a randomized uniform channel access decision by the SNs is due to their decentralized operation and 
lack of coordination between them. However, other channel access schemes can be accommodated by defining, more generally, the PMF $p_{R|T}(r|t),\ r\in\{0,1,\dots,t\}$.
\end{remark}

For both schemes, given the instruction policy $\delta$,
the sequence $\{(V_k,\pi_{\boldsymbol{\gamma},k}),k{\geq}0\}$ is a Markov chain.
In fact, the instruction $\mathbf D_k$ is chosen according to
$\delta_k(\mathbf D_{k}|V_k,\pi_{\boldsymbol\gamma,k})$.
Each SN decides its action $(A_{n,k},S_{M,n,k},B_{n,k})$ based on
$\mathbf D_k$ and $\gamma_{n,k}$, so that the aggregate SNR collected at the FC, $\Lambda_k$, is a random variable
 which only depends on $\mathbf D_k$ and $\boldsymbol{\gamma}_{k}$ and is independent of the past.
Finally, given $\Lambda_k$, from (\ref{nu}) and (\ref{nu2}) the next prior variance state is
$V_{k+1}{=}\nu(\hat \nu(V_{k},\Lambda_k))$, and $\boldsymbol{\gamma}_{k+1}$
only depends on $\boldsymbol{\gamma}_{k}$, whose distribution is $\pi_{\boldsymbol{\gamma},k}$, and is independent of other past events, so that the Markov property holds.
For the decentralized scheme, the next belief state $\pi_{\boldsymbol{\gamma},k+1}$ can be computed as a function of $\pi_{\boldsymbol{\gamma},k}$
 the measurements collected in slot $k$, and channel collisions.
On the other hand, for the coordinated scheme, $\pi_{\boldsymbol{\gamma},k+1}$ is a function of $\boldsymbol{\gamma}_{k+1}$,
whose value is fed back by the SNs.
\vspace{-3mm}
\begin{remark}
\label{remiid}
If $\alpha{=}0$, the process $X_k$ is i.i.d., hence $V_k{=}1,\ \forall k$. In this case, both schemes do not adapt to the quality feedback $V_k$, but only to the belief on the accuracy state 
$\pi_{\boldsymbol{\gamma},k}$. On the other hand, in the time-correlated case $\alpha\in (0,1)$, adaptation to the quality state $V_k$ may be necessary to achieve optimality, \emph{e.g.}, by instructing the SNs to remain idle if the quality of the estimate is good enough.
\end{remark}
\vspace{-8mm}
\subsection{Performance metrics and optimization problem}
\label{sec:optprob}
\noindent Given the initial value of the prior variance $V_0$, the initial distribution $\pi_{\boldsymbol\gamma,0}$, and the instruction policy $\delta$,
we define the average MSE and sensing-transmission cost of SN $n$  over a finite horizon of length $T+1$
as 
\begin{align}
\label{Cest}
&\bar M_{\delta}^{T}(V_0,\pi_{\boldsymbol\gamma,0})=\mathbb E\left[\left.R_T(V_0;\Lambda_0^T)\right|
V_0,\pi_{\boldsymbol\gamma,0}\right],
\\\label{CSN}
& {\bar C_{\delta}^{T,n}(V_0,\pi_{\boldsymbol\gamma,0})=\mathbb E\left[\left.
C_n^{T}(A_{n,0}^T,S_{M,n,0}^T)
\right|V_0,\pi_{\boldsymbol\gamma,0}\right],}
\end{align}
{
where $R_T(V_0;\Lambda_0^T)$ is the sample average MSE given by (\ref{RT}),
and $C_n^{T}(A_{n,0}^T,S_{M,n,0}^T)$
is the \emph{sample average sensing-transmission cost for SN $n$}, given by (\ref{ctn})}.
The expectation is computed with respect to the activation, local measurement SNR, accuracy state and medium access processes 
$\{\mathbf D_k,A_{n,k},S_{M,n,k},\gamma_{n,k},O_{n,k},n\in\{1,2,\dots,N_S\},k\in\mathbb N\}$, induced by policy $\delta$.
The goal is to determine $\delta^*$ such that
\begin{align}
\label{optprob}
\!\!\!\!\delta^*\!\!=\!\arg\min_{\delta} \bar M_{\delta}^{T}(V_0,\pi_{\boldsymbol\gamma,0}),
\text{s.t. }\bar C_{\delta}^{T,n}(V_0,\pi_{\boldsymbol\gamma,0})\!\leq\!\frac{\epsilon}{N_S},\forall n,\!\!
\end{align}
where $\epsilon{>}0$ is the maximum network cost constraint.
Alternatively,
we consider the Lagrangian formulation
\begin{align}
\label{optproblag}
\!\!\!\!\delta^*=&\arg\min_{\delta} \bar M_{\delta}^{T}(V_0,\pi_{\boldsymbol\gamma,0})
+\frac{\lambda}{c_{\mathrm{TX}}}\sum_{n=1}^{N_S}\bar C_{\delta}^{T,n}(V_0,\pi_{\boldsymbol\gamma,0}),
\end{align}
where $\lambda>0$ is the Lagrange multiplier, which trades off MSE and sensing-transmission cost.
 In particular, we are interested in the infinite horizon $T{\to}\infty$  (average long-term)
 and $V_0{=}1$,
 so that we will drop the dependence  on $T$ and $V_0$ in the following treatment, whenever possible.
 By varying $\epsilon$ in (\ref{optprob}) (respectively, $\lambda$ in (\ref{optproblag})),
we obtain different operational cost-MSE points $(\bar C_{\delta}^{n}(\pi_{\boldsymbol\gamma,0}),\bar M_{\delta}(\pi_{\boldsymbol\gamma,0}))$.
\vspace{-3mm}
\begin{remark}
\label{remoutage}
Note that the posterior variance process $\{\hat V_k,k{\geq}0\}$ may exhibit significant fluctuations over time, which may be undesirable. 
These fluctuations can be reduced by imposing a constraint on the frequency that a given MSE threshold $\hat v_{\mathrm{th}}$ is overcome,
defined by the \emph{outage} event $\hat V_k\geq\hat v_{\mathrm{th}}$, and by the time average expected outage
\begin{align}
\bar O_{\delta}^{T}(V_0,\pi_{\boldsymbol\gamma,0})=\frac{1}{T+1}
\mathbb E\left[\left.\sum_{k=0}^T\chi(\hat V_k\geq\hat v_{\mathrm{th}})\right|V_0,\pi_{\boldsymbol\gamma,0}\right].
\end{align}
The constraint $\bar O_{\delta}^{T}(V_0,\pi_{\boldsymbol\gamma,0}){\leq}\sigma$
can then be added to the optimization problem (\ref{optprob}), or the Lagrangian term $\mu \bar O_{\delta}^{T}(V_0,\pi_{\boldsymbol\gamma,0})$
to  (\ref{optproblag}).
The following DP algorithm (\ref{DPgen}) can be straightforwardly extended to this case. Its analysis is left for future~work.
\end{remark}
\vspace{-0.3cm}
\section{Analysis}
\label{analysis}
\noindent For the finite horizon $T{<}\infty$, for both the coordinated and decentralized schemes, the optimal instruction policy $\delta^*$,
which is the solution of (\ref{optproblag}), can be found
via DP \cite{Bertsekas2005}, by solving recursively, backward in time from $k=T$ to $k=0$,
 \begin{align}
\label{DPgen}
&\bar W^{T-k}(V_{k},\pi_{\boldsymbol\gamma,k})
=\min_{\delta_k(\cdot)}
\mathbb E\left[\left.
\bar W^{T-k-1}(V_{k+1},\pi_{\boldsymbol\gamma,k+1})\right|\delta_k\right]\nonumber
\\&
+\mathbb E\left[\left.
\hat \nu(V_{k},\Lambda_k)+
\frac{\lambda}{c_{\mathrm{TX}}}\sum_{n=1}^{N_S}c_{SN}(A_{n,k},S_{M,n,k})\right|\delta_k\right],
\end{align}
where $V_{k+1}{=}\nu(\hat \nu(V_{k},\Lambda_{k}))$ and $\bar W^{-1}(V_{T+1},\pi_{\boldsymbol\gamma,T+1}){=}0$.
The minimizer is the optimal instruction policy $\delta_k^*(\cdot)$ in slot $k$,
and $\bar W^{T}(V_0,\pi_{\boldsymbol\gamma,0})/(T+1)$ yields the optimal cost function for the Lagrangian problem (\ref{optproblag}).
The infinite horizon scenario $T\to\infty$ can be approximated by choosing $T$ sufficiently large.
In general, (\ref{DPgen}) has high complexity, due to the large action space, non-convex nature, and the dependence on the accuracy state belief 
$\pi_{\boldsymbol\gamma,k}$.
In particular, in the coordinated scheme, the optimization is over the joint action $(A_{n,k},S_{M,n,k},B_{n,k})$
of each SN, as a function of $V_k$ and $\boldsymbol{\gamma}_k$ and time $k$.
On the other hand, in the decentralized scheme, the optimization is over functions
$q_k{:}\Gamma{\mapsto}[0,1]$ and $S_{M,k}{:}\Gamma{\mapsto}[0,\infty)$.
To overcome these dimensionality issues,
in Secs.~\ref{centr} and \ref{deccentr} we derive structural properties of the optimal policy and of the cost function by exploiting the \emph{statistical symmetry}
 and the \emph{large network} approximation $N_S{\gg}1$,
which enable a more efficient solution of (\ref{DPgen}). In Part II, we will further reduce the complexity by proposing
 near-optimal myopic policies.
Theorem \ref{lowbound} lower bounds the optimal MSE under any scheme.
\begin{thm}
\label{lowbound}
If $T=\infty$, we have
$\bar M_{\delta^*}\!\!\geq\!\!\hat\nu^*(\bar \Lambda^*)$,
where
\begin{align}
\label{nuing}
&\!\!\hat\nu^*(x)\triangleq\frac{
\sqrt{\!(1\!-\!\alpha)^2(1\!+\!x^2)\!+\!2(1\!\!-\!\!\alpha^2)x}
\!-\!(1\!-\!\alpha)(1\!+\!x)
}
{
2\alpha x
},
\\
\label{maxagg}
&\!\!\!\!\bar\Lambda^*\!\!=\!\max_{\delta}\mathbb E\left[\left.\Lambda_k\right|\delta\right]\!,\text{s.t.}
\mathbb E\left[\left.c_{SN}(A_{n,k},S_{M,n,k})\right|\delta\right]\!\!\leq\!\!\frac{\epsilon}{N_S},\!\forall n.\!\!\!
\end{align}
\end{thm}
\noindent\emph{Proof:}
The proof follows from the fact that $R_T(1;\Lambda_0^T)$ is a convex function of $\Lambda_0^T$ (Prop.~\ref{propV} in App.~\ref{appprop}),
hence $\bar M_{\delta^*}^T{\geq}R_T(1;\mathbb E\left[\left.\Lambda_0^T\right|\delta^*\right])$. 
Letting $\bar \Lambda{=}\frac{1}{T+1}\sum_{k}\mathbb E\left[\left.\Lambda_k\right|\delta^*\right]$ be the average aggregate SNR, we have 
$R_T(1;\mathbb E\left[\left.\Lambda_0^T\right|\delta^*\right]){\geq}R_T^*(\bar\Lambda)$,
where $R_T^*(\bar\Lambda)$ is defined in (\ref{RTstar}). Since $R_T^*(x)$ is a decreasing function of $x$ (Theorem~\ref{thm3} in App.~\ref{appprop}) and 
$\bar\Lambda{\leq}\bar\Lambda^*$ as a result of the optimization in (\ref{maxagg}),
we also have $R_T^*(\bar\Lambda){\geq}R_T^*(\bar\Lambda^*)$. Finally, in the limit $T{\to}\infty$, using Corollary~\ref{corollary} in App.~\ref{appprop},
we obtain $\bar M_{\delta^*}{\geq}\underset{T\to\infty}{\lim}R_T^*(\bar\Lambda^*){=}\hat \nu^*\left(\bar\Lambda^*\right)$, proving the theorem.\hfill\QED

The policy solving the optimization problem (\ref{maxagg}) is denoted as the \emph{max aggregate SNR scheme} (MAX-SNR). In each slot, it maximizes the expected 
aggregate SNR collected at the FC, under the cost constraint for the SNs.
MAX-SNR is non-adaptive, since it is independent of $V_k$.
The lower bound in Theorem \ref{lowbound} can be achieved only if the aggregate SNR $\Lambda_k{=}\bar\Lambda^*$ is collected deterministically in each slot
 (Corollary~\ref{corollary} in App.~\ref{appprop}).
However, this lower bound is, in general, not achievable, 
since the cross-layer factors introduce uncertainties and random fluctuations of the aggregate SNR $\Lambda_k$ around its mean, thus
degrading the MSE performance. Hence, MAX-SNR may achieve poor performance in general, as shown in Sec. \ref{numres}.
We now analyze both schemes.
\vspace{-3mm}
\subsection{Analysis  of Coordinated scheme}
\label{centr}
\noindent In the coordinated scheme, collisions can be avoided by scheduling at most one SN to transmit in each channel.
Without loss of optimality, the SNs are scheduled to transmit, in order, in
the channels with ID $1,2,\dots B$. Therefore,
if $A_{n,k}{=}1$, we let $B_{n,k}{=}\sum_{m=1}^{n}A_{m,k}$. This channel scheduling is optimal, since the $B$ orthogonal channels are symmetric and interchangeable.
We proceed as follows. We first derive structural properties of the optimal policy and of the DP
algorithm by exploiting the statistical symmetry of the WSN for the \emph{best-}$\gamma$ scenario, yielding a lower bound to the MSE
achievable under the  \emph{Markov-}$\gamma$ scenario.
Based on that, we then
design low-complexity policies for the
 \emph{Markov-}$\gamma$ scenario, 
 which are shown to be near-optimal for large WSNs.
 \subsubsection{\emph{Best-}$\gamma$ scenario}
In this case, the belief $\pi_{\boldsymbol{\gamma},k}$ is constant and  can be neglected.
 Prop. \ref{optsymm} states the optimality of \emph{policy symmetry}, \emph{i.e.}, due to the statistical symmetry of the WSN,
it is optimal for the FC to schedule actions \emph{uniformly} randomly  across SNs.
In other words, the SNs incur the same sensing-transmission cost
and have the same sensing capabilities, hence there is no preference of one SN over another.
Let
\begin{align*}
\mathcal D^{(O)}\equiv\{\mathbf D\in\mathcal D:A_{n}\geq A_{n+1},\forall n; S_{M,n}\geq S_{M,n+1},\forall n\}
\end{align*}
be an ordered subset of instructions. We have that any instruction $\mathbf D{\in}\mathcal D$ can be obtained
by permutation of some $\mathbf D^{(O)}{\in}\mathcal D^{(O)}$. Additionally, let $\mathcal D(\mathbf D^{(O)})$ be the subset of instructions in
$\mathcal D$ obtained by permutation of the entries of $\mathbf D^{(O)}$,
so that $\mathcal D{\equiv}\cup_{\mathbf D^{(O)}\in\mathcal D^{(O)}}\mathcal D(\mathbf D^{(O)})$. 
\begin{propos}
\label{optsymm}
In the \emph{best-}$\gamma$ scenario,
one optimal instruction policy $\delta^*$ for (\ref{optprob}) or (\ref{optproblag}) satisfies, $\forall V_k$,
\begin{align*}
\delta_k^*(\mathbf D|V_{k})= \delta_k^*(\mathbf D^{(O)}|V_{k}),\ \forall \mathbf D\in \mathcal D(\mathbf D^{(O)}),\ 
\forall \mathbf D^{(O)}\in\mathcal D^{(O)}.
\end{align*}
\end{propos}
\noindent\emph{Proof:}
See App.~\ref{Proofoptsymm}.
\hfill\QED

\noindent 
We denote an instruction policy satisfying the hypothesis of Prop.~\ref{optsymm}
as a \emph{symmetric instruction policy}. Such a policy is symmetric with respect to the SN scheduling, and induces
 the same expected cost for each SN, so that the superscript $n$ in (\ref{CSN}) can be neglected.
To generate a symmetric instruction policy,
the FC first selects
 one ordered instruction $\mathbf D^{(O)}$ from the lower-dimensional set $\mathcal D^{(O)}$, and then
  assigns, in order, each component of $\mathbf D^{(O)}=(d_1^{(O)},d_2^{(O)},\dots,d_{N_S}^{(O)})$ to a random SN, until
all of them have been scheduled. 
The following proposition
  demonstrates the optimality of
allocating the same local measurement SNR to all active SNs.
\begin{propos}
\label{lem2}
In the \emph{best-}$\gamma$ scenario,
the optimal $\delta^*$ allocates $S_{M,n,k}=S_{M,k}$ for all $n$ such that $A_{n,k}=1$.
\end{propos}
\noindent\emph{Proof:}
See App.~\ref{Proofoptsymm}.
\hfill\QED

\noindent
This result follows from the concavity of the aggregate SNR with respect to the SNR allocation of the SNs.
Under the resource constraints, it is thus optimal for the SNs to employ the same SNR, in order to maximize the aggregate SNR collected at the FC.
 From Props. \ref{optsymm} and \ref{lem2}, it follows that, in the \emph{best-}$\gamma$ scenario, it is sufficient for the FC to choose, in each slot $k$,
the number of SNs to activate $t_k\in\{0,1,\dots,B\}$, and their common local measurement SNR $S_{M,k}$. The $t_k$ active SNs
are then chosen uniformly from the set of SNs.
For a given pair $(t_k,S_{M,k})$,
the aggregate SNR collected at the FC is thus $\Lambda_k=t_k\frac{S_AS_{M,k}}{S_A+S_{M,k}}$.
The MSE performance is governed by the aggregate SNR $\Lambda_k$ collected at the FC. Since the FC 
can control $(t_k,S_{M,k})$, we can optimize these two quantities to minimize the sensing-transmission cost
in order to collect the target aggregate SNR $\Lambda_k$ at the FC,
denoted as $(t^*(\Lambda_k),S_{M}^*(\Lambda_k))$,
 yielding the following proposition.
\begin{propos}
\label{lem3}
Let $\Lambda{<}BS_A$ be the target aggregate SNR collected at the FC.
In the \emph{best-}$\gamma$ scenario,
if $\Lambda{=}0$, then $t^*(0){=}0$ and $S_{M}^*(0)=0$.
Otherwise ($\Lambda>0$), let
\begin{align*}
\Lambda_{\mathrm{th}}(t)\triangleq\frac{2S_At(t+1)}
{\sqrt{1+4S_A\theta t(t+1)}+2t+1};
\end{align*}
then $t^*(\Lambda){=}\min\{t,B\}$
and $S_{M}^*(\Lambda)=\frac{S_A\Lambda}{t^*(\Lambda)S_A-\Lambda}$, where $t{\geq}1$ is the unique value such that $\Lambda\in[\Lambda_{\mathrm{th}}(t-1),\Lambda_{\mathrm{th}}(t))$.
\end{propos}
\noindent\emph{Proof:}
See App.~\ref{Proofoptsymm}.
\hfill\QED

\noindent From Prop.~\ref{lem3}, it follows that it is sufficient for the FC to determine, in each slot $k$,
the target aggregate SNR $\Lambda_k$. The number of SNs activated is then given by $t_k=t^*(\Lambda_k)$,
 and the common local measurement SNR is $S_{M,n,k}=S_{M}^*(\Lambda_k)$.
  Note that $\Lambda_{\mathrm{th}}(t)$ is an increasing function of $t$, implying that an increasing number of SNs need to be activated 
as the aggregate SNR requirement $\Lambda_k$ increases.
Moreover, $\Lambda_{\mathrm{th}}(t)$ is an increasing function of $S_A$ 
and decreasing function of the normalized unitary sensing cost $\theta$, so that, as $S_A$ grows or $\theta$ diminishes,
less SNs need to be activated.
In fact, $S_A$ determines the error floor in the measurement collected by each SN. 
Therefore, as $S_A$ increases and the ambient noise becomes less relevant, it is sufficient to activate 
a smaller number of SNs with higher SNR, in order to reduce the transmission cost.
Similarly, as $\theta$ grows, the
transmission cost becomes less and less relevant with respect to the sensing cost, hence
more SNs can be activated.
We thus obtain:

\noindent {\bf COORD-DP: DP algorithm for the coordinated scheme, \emph{best-}$\gamma$ scenario.}
 For $k=T,T-1,\dots,0$, solve, $\forall V_k{\in}[1-\alpha,1]$,
  \begin{align}\label{DPgenCOORD}
  \nn
&\bar W^{T-k}(V_{k})
=\!\!\!\!\min_{\Lambda_k\in [0,BS_A)}
\bar W^{T-k-1}(\nu(\hat \nu(V_{k},\Lambda_{k})))
\\&
+\hat \nu(V_{k},\Lambda_k)
+\frac{\lambda}{c_{\mathrm{TX}}} t^*(\Lambda_k)c_{SN}\left(1,S_{M}^*(\Lambda_k)\right),
\end{align}
where $\bar W^{-1}(V_{T+1})=0$. 
The optimizer, $\Lambda_k^*(V_k)$, is the optimal aggregate SNR collected at the FC in slot $k$.
\hfill\QED

Note that, by exploiting the statistical symmetry of the WSN, we have enabled a significant complexity reduction with respect to 
 (\ref{DPgen}), since the optimization is only over the aggregate SNR sequence, rather than the joint action $(A_{n,k},S_{M,n,k},B_{n,k})$ of each SN.
 
The next theorem characterizes regimes of $\epsilon$ where the optimal policy is the MAX-SNR scheme.
   \begin{thm}
 \label{thm1}
In the \emph{best-}$\gamma$ scenario with $T\to\infty$,
\\(i) if $\epsilon=tc_{\mathrm{TX}}(1+\sqrt{\theta S_A})$, for some $t=1,2,\dots,B$,
 $\delta^*$ is the MAX-SNR scheme, 
with $t_k^*=t$, $S_{M,k}^*=\sqrt{\frac{S_A}{\theta}},\ \forall k$;
\\(ii) if $\epsilon>Bc_{\mathrm{TX}}(1+\sqrt{\theta S_A})$,
 $\delta^*$ is the MAX-SNR scheme, 
with $t_k^*=B$, $S_{M,k}^*=\frac{1}{\theta}\left(\frac{\epsilon}{Bc_{\mathrm{TX}}}-1\right),\ \forall k$;
\\(iii) in both cases,  $\bar M_{\delta^*}=\hat\nu^*\left(t_k^*\frac{S_AS_{M,k}^*}{S_A+S_{M,k}^*}\right)$, where $\hat\nu^*(x)$ is given by (\ref{nuing}),
and $\bar C_{\delta^*}=\frac{\epsilon}{N_S}$.
\end{thm}
\noindent\emph{Proof:}
See App.~\ref{Proofthm1}.
\hfill\QED

Theorem \ref{thm1} follows from the fact that, in the \emph{best-}$\gamma$ scenario, the FC can deterministically control the quality of the measurements 
collected in each slot (aggregate SNR $\Lambda_k$), \emph{i.e.}, there are no uncertainties. If the condition on
 $\epsilon$ given by Theorem~\ref{thm1} is satisfied, the FC can thus schedule each SN so as to collect a constant aggregate SNR (the highest possible, under the resource constraints, as dictated by the MAX-SNR scheme), thus achieving the lower bound in Theorem~\ref{lowbound} (see comments therein).
 On the other hand, if the condition on $\epsilon$ is not satisfied, the FC may need to resort to time-sharing in order to best exploit all available resources. The policy
  in this case can be obtained via the DP in (\ref{DPgenCOORD}).
 
 Note that, in case (i), the local measurement SNR $S_{M,k}^*$ only depends on $S_A$ and $\theta$.
 In particular, it is an increasing function of $S_A$ and decreasing function of $\theta$.
 In fact, if $S_A$ increases, the error floor represented by the ambient noise diminishes, hence
 more accurate measurements can be collected; similarly, if $\theta$ increases, sensing becomes more costly, hence
 $S_{M,k}^*$ diminishes.
 On the other hand, in case (ii), sensing-transmission resources are abundant to SNs, hence $B$ SNs are activated in order to saturate all $B$ channels.
 $S_{M,k}^*$ in this case is selected in such a way as to use up all available resources.

 \subsubsection{\emph{Markov-}$\gamma$ scenario}
 \label{MarkovCOORD}
 In this case, $\gamma_{n,k}$ fluctuates over time, thus causing random fluctuations
in the aggregate SNR collected at the FC.
 The optimal policy is difficult to characterize,
 due to the high dimensionality of the problem.
 Herein, we define a sub-optimal policy, based 
 on the optimal DP policy derived in the previous section.
 To this end, 
 let $r(\cdot;\boldsymbol{\gamma}_k):\{1,2,\dots,N_S\}\mapsto\{1,2,\dots,N_S\}$
be a ranking of SNs indexed by $\boldsymbol{\gamma}_k$, such that $r(m;\boldsymbol{\gamma}_k)$
is the label of the SN with the $m$th highest accuracy state,
\emph{i.e.}, $\gamma_{r(1;\boldsymbol{\gamma}_k),k}\geq\gamma_{r(2;\boldsymbol{\gamma}_k),k}\geq,\dots,\geq\gamma_{r(N_S;\boldsymbol{\gamma}_k),k}$. 
  Let $\delta^*$ be the optimal policy solving
(\ref{optprob}) or (\ref{optproblag}) for the \emph{best-}$\gamma$ scenario,
 $\{t_k^*,S_{M,k}^*,\Lambda_k^*,k\geq 0\}$ be the sequence of number of active SNs, local measurement and aggregate SNRs generated by such policy
in the \emph{best-}$\gamma$ scenario.
Denote the optimal MSE and cost in the \emph{best-}$\gamma$ scenario as
$\bar M_{\delta^*}^{\gamma_{\max}}$ and $\bar C_{\delta^*}^{\gamma_{\max}}$, respectively.
Clearly, $\Lambda_k^*$ is an upper bound to the aggregate SNR collected at the FC in the 
\emph{Markov-}$\gamma$ scenario, due to the fluctuations in the local accuracy state.
 Let $\{\tilde V_k,k{\geq}0\}$ be a \emph{virtual prior variance process}, obtained as if all measurements were collected with
 the best accuracy state $\gamma_{\max}$. Starting from $\tilde V_0=V_0$, this can be generated recursively as $\tilde V_{k+1}=\nu(\hat\nu(\tilde V_k,\Lambda_k^*))$.
We define the \emph{sub-optimal coordinated DP policy} (SCDP) as follows.

\noindent
\textbf{SCDP}:
Given $(\tilde V_k,\boldsymbol\gamma_k)$, SCDP allocates the $t_k^{*}$  SNs with the best accuracy state,
 with local measurement SNR $S_{M,k}^{*}$,
\begin{align}
\!\!\nonumber
\left\{
\begin{array}{l}
\!\!\!\!A_{r(m;\boldsymbol{\gamma}_k),k}\!=\!1,\ \!\!S_{M,r(m;\boldsymbol{\gamma}_k),k}\!=\!S_{M,k}^*,\forall m\!=\!1,2,\!\dots\!,\!t_k^*,
\\
\!\!\!\!A_{r(m;\boldsymbol{\gamma}_k),k}\!=\!0,\ \forall m>t_k^*.\hfill\QED
\end{array}
\right.
\end{align}

We have the following theorem.
\begin{thm}
\label{thm2}
In the \emph{Markov-}$\gamma$ scenario, if $N_S\geq \frac{B-1}{\pi_{\gamma}(\gamma_{\max})}$, 
under SCDP,
 $\!\bar C_{\delta^*}\!=\!\bar C_{\delta^*}^{\gamma_{\max}}\!$ and
\begin{align}
&
\!\!\!0\!\leq\!\bar M_{\delta^*}\!\!-\!\bar M_{\delta^*}^{\gamma_{\max}}
\!\!\leq\!
\frac{1}{1\!-\!\alpha}\!\exp\!\left\{\!\!-\frac{\left(N_S\pi_{\gamma}(\gamma_{\max})\!-\!B\!+\!1\right)^2}{2N_S\pi_{\gamma}(\gamma_{\max})}\!\!\right\}\!.\!\!\!
\label{inex}
\end{align}
\end{thm}
\noindent\emph{Proof:}
See App.~\ref{Proofthm2}.
\hfill\QED

\noindent{Theorem \ref{thm2} states that SCDP
 achieves the same sensing-transmission cost as the optimal policy in the  \emph{best-}$\gamma$ scenario.
 This is by construction and due to the statistical symmetry property, since all SNs experience the same steady-state distribution of their local accuracy state,
 hence each of them belongs to the set of $t_k^*$ best SNs with the same frequency.
 On the other hand, the MSE gap with respect to the lower bound represented by the optimal policy in the  \emph{best-}$\gamma$ scenario
 decreases exponentially with the network size $N_S$.
Therefore,  SCDP is nearly optimal for $N_S$ sufficiently large.}
Alternatively,
a densely deployed WSN
 provides \emph{sensing diversity}, \emph{i.e.}, in each slot, a sufficiently large pool of SNs
can sense $X_k$ with high accuracy, despite the fluctuations in the \emph{local} accuracy state of each SN.
SCDP can be optimized efficiently via
the DP in (\ref{DPgenCOORD}) for the \emph{best-}$\gamma$ scenario,
and is given by Theorem~\ref{thm1}, if the condition on $\epsilon$ holds.
\vspace{-4mm}
\subsection{Analysis of Decentralized scheme}
\label{deccentr}
\noindent In this section, we analyze the decentralized scheme.
By adapting the DP in (\ref{DPgen}) to this case, we obtain
\begin{align}
\label{DP}
&\bar W^{T-k}(V_{k},\pi_{\boldsymbol\gamma,k})
=
\min_{\!\!\!\!\!\!\!(q,S_M):\Gamma\mapsto[0,1]\times[0,\infty)\!\!\!\!\!\!\!}
\mathbb E\left[\left.\hat \nu\left(V_{k},\Lambda_k\right)\right|q,S_M\right]
\nonumber\\&
+
\frac{\lambda}{c_{\mathrm{TX}}} \!\!\!\sum_{\boldsymbol\gamma\in\Gamma^{N_S}}\pi_{\boldsymbol\gamma,k}(\boldsymbol\gamma)\sum_{n=1}^{N_S}q(\gamma_n) c_{SN}(1,S_{M}(\gamma_n))
\nonumber
\\&
+
\mathbb E\left[\left.
\bar W^{T-k-1}\left(\nu\left(\hat \nu\left(V_{k},\Lambda_k\right)\right),\pi_{\boldsymbol\gamma,k+1}\right)\right|q,S_M\right],
\end{align}
where
$\Lambda_k{=}\sum_{n}O_{n,k}\gamma_{n}^2S_AS_{M}(\gamma_{n})/(S_A{+}S_{M}(\gamma_{n}))$,
whose distribution depends on $q(\cdot)$ and $S_M(\cdot)$ via $(O_{n,k},\gamma_n),\forall n$.
As in the coordinated scheme, we first study the \emph{best}{-}$\gamma$ scenario, and then
 extend our analysis to the \emph{Markov-}$\gamma$ scenario.
\subsubsection{\emph{Best-}$\gamma$ scenario}
\label{bestomegadistr}
Letting $\gamma_n{=}1,\forall n$, we obtain $\Lambda_k{=}\frac{R_kS_AS_{M}}{S_A+S_{M}}$,
where $R_k{=}\sum_{n=1}^{N_S}O_{n,k}$ is the number of packets successfully received at 
the FC, with PMF $p_R(R_k;q)$.
\begin{propos}
\label{psr}
If the SNs activate with probability~$q$,
 \begin{align*}
p_R(r;q)\!\!=\!\!
\sum_{k=r}^{B}\!
\frac{(-1)^{k-r}N_S!}{(N_S-k)!}\!\!
\left(\begin{array}{c}\!\!\!\!B\!\!\!\!\\\!\!\!\!r\!\!\!\!\end{array}\right)\!\!
\left(\begin{array}{c}\!\!\!B-r\!\!\!\\\!\!\!k-r\!\!\!\end{array}\right)\!\!
\left(\frac{q}{B}\right)^{k}\!\!
\left(1\!-\!k\frac{q}{B}\right)^{\!\!N_S-k}\!\!\!\!\!\!\!.
\end{align*}
\end{propos}
\noindent\emph{Proof:}
See App.~\ref{proofoflemchannel}.
\hfill\QED

\noindent We employ the large network approximation $N_S{\gg}1$ to approximate $p_R(r;q)$.
We define the \emph{normalized activation probability per channel},
$\zeta=qN_S/B$, and let $N_S\to\infty$ with $\zeta$ fixed. 
We thus obtain the following corollary of Prop.~\ref{psr}.
\begin{corol}
\label{corol1}
When $N_S\to\infty$,
$R_k$
 has binomial distribution with $B$ trials and success probability $\zeta e^{-\zeta}$
in each channel, denoted as $p_R(R_k;\zeta)$.
\end{corol}

\noindent {The implication is that the successes/collisions are independent across channels, each Bernoulli distributed. 
This is not true for finite $N_S$, since the transmissions are coupled (each active SN transmits on a unique channel), hence
the successes/collisions are correlated across channels,
but it enables a good tractable approximation for finite $N_S$.}
Using the large network approximation, DP is given as follows.

\noindent {\bf DEC-DP: DP algorithm for the decentralized scheme, \emph{best-}$\gamma$ scenario.}
 For $k=T,T-1,\dots,0$, solve, $\forall V_k{\in}[1-\alpha,1]$,
  \begin{align}
  \label{DPzeta}
&\!\!\!\!\bar W^{T-k}(V_{k})
\!\!=\!\!\min_{\zeta,S_{M}}\!\sum_{r=0}^Bp_R(r;\zeta)\hat \nu\left(\!\!V_{k},r\frac{S_AS_M}{S_A\!\!+\!\!S_M}\!\right)
\!\!+\!\!\frac{\!\lambda\zeta\!}{\!c_{\mathrm{TX}}\!} c_{SN}(1,\!S_{M}\!)
\nonumber
\\&
\!+\!\sum_{r=0}^Bp_R(r;\zeta)
\bar W^{T-k-1}\left(\nu\left(\hat \nu\left(V_{k},r\frac{S_AS_M}{S_A+S_M}\right)\right)\right),
\end{align}
where $\bar W^{-1}(V_{T+1})=0$. 
The optimizer, $(\zeta_k^*(V_k),S_{M,k}^*(V_k))$, is the optimal normalized activation probability and local measurement SNR pair in slot $k$.
\hfill\QED

The activation probability when $N_S<\infty$ can then be approximated by $q_{k}^*(V_k)\simeq\zeta_k^*(V_k) B/N_S$.
Due to the shared wireless channel, the transmission probability of the SNs should be bounded, as stated in the following proposition.
\begin{propos}
\label{zetaopt}
When $N_S\to\infty$, the normalized transmission probability per channel satisfies $\zeta_k^*(V_k)\leq 1$.
\end{propos}
\noindent\emph{Proof:}
See App.~\ref{proofoflemchannel}.
\hfill\QED
\vspace{-0.2cm}
\begin{remark}
\label{rem2}
Note that, if $B{=}1$, the success rate $N_Sq(1{-}q)^{N_S-1}$ is maximized by $q{=}1/N_S$, \emph{i.e.}, $\zeta{=}1$.
Any $q{>}1/N_S$ ($\zeta{>}1$) incurs higher cost and collision probability, hence worse MSE performance, and is thus sub-optimal.
Therefore, Prop.~\ref{zetaopt} holds trivially for $B{=}1,\forall N_S$.
For $B>1$, this result holds for $N_S\to\infty$, since channel outcomes are decoupled in this case (Corollary~\ref{corol1}).
\end{remark}
From Prop.~\ref{zetaopt}, the minimization in the DP stage (\ref{DPzeta})
can be confined to $\zeta\in [0,1]$, thus reducing the search space.

\subsubsection{\emph{Markov-}$\gamma$ scenario}
 \label{MarkovDIST}
 The optimal policy for this case is difficult to characterize, due to the high dimensionality of the problem.
 Similar to the coordinated scheme, we define the following \emph{sub-optimal decentralized DP policy} (SDDP).
 To this end, let $(\zeta_k^*(V_k),S_{M,k}^*(V_k))$ be the optimal policy under the \emph{best-}$\gamma$ scenario, obtained via (\ref{DPzeta}).
 
 \noindent\textbf{SDDP}:
Given $V_k$,
the activation probability is defined as
\begin{align}
q_k(V_k,\gamma)=
\left\{
\begin{array}{ll}
1,&\gamma >\gamma_{\mathrm{th}},\\
\frac{\frac{B}{N_S}\zeta_k^{*}(V_k)-\sum_{\gamma>\gamma_{\mathrm{th}}}\pi_{\gamma}(\gamma)}{\pi_{\gamma}(\gamma_{\mathrm{th}})},
&
\gamma=\gamma_{\mathrm{th}},\\
0,&\gamma<\gamma_{\mathrm{th}},
\end{array}
\right.
\end{align}
and the local measurement SNR as $S_{M,n,k}\!{=}S_{M,k}^*\!(V_k)$,
 where $\gamma_{\mathrm{th}}{\in}\Gamma$ uniquely solves
$\!\!\!\underset{\gamma\geq \gamma_{\mathrm{th}}}{\sum}\!\!\!\!\pi_{\gamma}(\gamma){\geq}
B\zeta_k^{*}(V_k)/N_S{>}\!\!\!\!\underset{\gamma{>}\gamma_{\mathrm{th}}}\sum\!\!\!\pi_{\gamma}(\gamma)$.\hfill\QED

\noindent Note that, under SDDP,
$\sum_{\gamma}q_k(V_k,\gamma)\pi_{\gamma}(\gamma)N_S/B{=}\zeta_k^{*}(V_k)$,
\emph{i.e.}, each SN activates with \emph{marginal} normalized probability $\zeta_k^{*}(V_k)$,
with respect to the steady-state distribution of $\gamma_{n,k}$.
For the \emph{i.i.d.-}$\gamma$ scenario, we have the following proposition.
\vspace{-3mm}
\begin{propos}
SDDP is optimal in the \emph{i.i.d.-}$\gamma$ scenario,
if $N_S{\geq}\frac{B}{\pi_\gamma(\gamma_{\max})}$.
\end{propos}
As in the coordinated scheme, this result is a consequence of the fact that a densely deployed WSN
 provides \emph{sensing diversity}, \emph{i.e.}, in each slot, a sufficiently large pool of SNs
can sense the underlying process with high accuracy, despite the fluctuations in the \emph{local} accuracy state of each SN.
In particular, if $N_S{\geq}B/\pi_\gamma(\gamma_{\max})$,
then SDDP yields $q_k(V_k,\gamma_{\max})=\frac{B\zeta_k^{*}(V_k)}{N_S\pi_{\gamma}(\gamma_{\max})}$,
$q_k(V_k,\gamma)=0,\forall\gamma<\gamma_{\max}$, so that only the SNs with the best accuracy state may activate, and no loss is incurred with respect to
the \emph{best-}$\gamma$ scenario.
On the other hand, if $N_S{<}B/\pi_\gamma(\gamma_{\max})$,
the FC may resort to the SNs with lower accuracy to sense and report their measurement.
The DP (\ref{DP}) for the general case has high complexity, due
 to the high-dimensional action space (the activation probability and local measurement SNR are functions of the accuracy state) and state space
(the belief $\pi_{\boldsymbol\gamma,k}$ is part of the state).
 Moreover, the optimal DP policy in the \emph{Markov-}$\gamma$ scenario is cumbersome  to operate, since the FC needs to track the 
belief $\pi_{\boldsymbol\gamma,k}$.
In contrast, SDDP has lower optimization and operational complexity, since it is optimized for the \emph{best-}$\gamma$ scenario and
it does not require the FC to track $\pi_{\boldsymbol\gamma,k}$.
\vspace{-0.3cm}
\subsection{Cost of communication overhead}
\label{commover}
\noindent In this section, we evaluate the communication overhead required to implement the two schemes,
assuming the sub-optimal DP policy is used in the \emph{Markov-}$\gamma$ scenario.
In the \emph{uplink} channel (SNs to FC), each SN incurs the cost $c_{\gamma}$ to report its accuracy state to the FC. 
On the other hand, in the
\emph{downlink} channel (FC to SNs), the FC incurs the cost $c_{V}$ to feed back the quality state $V_k$, and $c_{SC}$ to schedule each SN to activate. 
The mapping of $V_k$ to the corresponding sensing-transmission action is stored in each SN in a look-up table.
\subsubsection{Coordinated scheme}
In this scheme, the SNs need to report their accuracy state, whenever it changes.
Therefore, the (average long-term) uplink communication overhead of the network is $\bar C_{UOH}{=}N_Sc_{\gamma}\sum_{\gamma\in\Gamma}\pi_{\gamma}(\gamma)(1{-}P_\gamma(\gamma;\gamma))
$, which grows with the WSN size.
In particular, $\bar C_{UOH}{=}0$ in the \emph{best-}$\gamma$ scenario and 
$\bar C_{UOH}{=}N_Sc_{\gamma}\left(1{-}\sum_{\gamma\in\Gamma}\pi_{\gamma}(\gamma)^2\right)$ in the \emph{i.i.d.-}$\gamma$ scenario;
in the general  \emph{Markov-}$\gamma$ scenario, $\bar C_{UOH}$ is small if $P_\gamma(\gamma;\gamma)\simeq 1,\ \forall \gamma$, \emph{i.e.}, the accuracy state varies slowly over time.
In the downlink channel, the FC schedules each SN individually, hence the downlink communication overhead is $t_k^*c_{SC}$ in slot $k$, since $t_k^*$ are scheduled to activate.
Since $t_k^*\leq B$, the average long-term downlink communication overhead satisfies $\bar C_{DOH}\leq Bc_{SC}$.
\subsubsection{Decentralized scheme}
In this scheme, the SNs do not report their local accuracy state to the FC, hence $\bar C_{UOH}=0$.
On the other hand, in the downlink channel, the FC broadcasts the quality state $V_k$ in each slot, hence $\bar C_{DOH}=c_V$.

 \begin{figure}[t]
\centering
\includegraphics[width = .85\linewidth,trim = 10mm 4mm 10mm 9mm,clip=true]{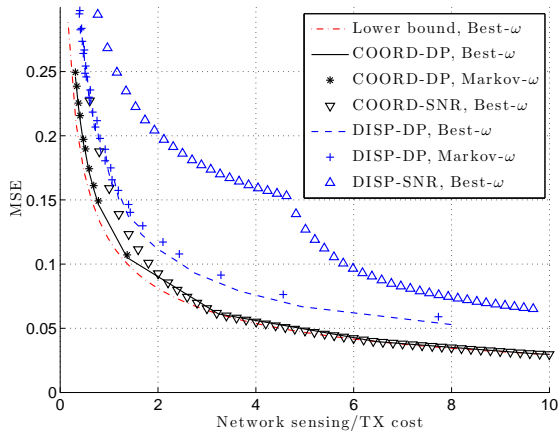}
\vspace{-3mm}
\caption{MSE as a function of the network cost, $N_S=20$.}
\label{NS100B1}
\vspace{-5mm}
\end{figure}

Note that, unlike the coordinated scheme, the decentralized one incurs no uplink communication overhead cost. If $Bc_{SC}>c_V$,
it incurs also a smaller downlink communication overhead cost. Therefore, overall, the decentralized scheme is more scalable to large WSNs. As we will see in the next section, 
this improved scalability and lower communication overhead come at the cost of MSE degradation.
\vspace{-3mm}
\section{Numerical Results}\label{numres}
\noindent In this section, we provide numerical results.
Unless otherwise stated, we consider a WSN of size $N_S{\in}\{20,100\}$ (\emph{small} and \emph{large} WSN, respectively).
We model $\{\gamma_{n,k}\}$
as a Markov chain taking values in the set $\Gamma{\equiv}\{\sqrt{i/10},i{=}1,2,\dots,10\}$, with transition probabilities
$P_\gamma(\gamma;\gamma){=}0.9$,
$P_\gamma(\sqrt{1/10};\sqrt{2/10}){=}P_\gamma(1;\sqrt{9/10}){=}0.1$,
$P_\gamma(\sqrt{i/10};\sqrt{(i+1)/10}){=}P_\gamma(\sqrt{i/10};\sqrt{(i-1)/10}){=}0.05$, $i{=}2,3,\dots,9$.
We let $c_{\mathrm{TX}}{=}1$,\footnote{{Note that the choice $c_{\mathrm{TX}}=1$ is without loss of generality, since, by scaling
$c_{\mathrm{TX}}$ and $\phi$ by the same value, while keeping the normalized unitary sensing cost $\theta$ constant,
 the long-term sensing-transmission cost scales accordingly, without changing
the form of the optimal policy, and without providing any further insights.}} $S_A=20$, $\phi=0.25$, $\alpha=0.96$, and $B=5$.
We consider the following schemes for the \emph{best-}$\gamma$ scenario:

\noindent
$\bullet$ \emph{COORD-DP}: coordinated scheme, obtained via the DP in (\ref{DPgenCOORD}) or given by
 Theorem~\ref{thm1}, if the condition on $\epsilon$ holds;
 
 \noindent$\bullet$  \emph{DEC-DP}: the decentralized scheme considered in Sec.~\ref{deccentr},
obtained via the DP in (\ref{DP});

\noindent
$\bullet$ \emph{COORD-SNR}: 
MAX-SNR policy for the coordinated scheme (see (\ref{opwreter})),
determined in the proof of Theorem~\ref{thm1} in App.~\ref{Proofthm1};

\noindent
$\bullet$ \emph{DEC-SNR}: 
MAX-SNR policy for the decentralized scheme,
\begin{align*}
(\zeta^*,S_M^*){=}\underset{\zeta,S_M}{\arg\max}\frac{\mathbb E[\left.R_k\right|\zeta]S_AS_M}{S_A+S_M}\ \text{s.t.}\ B\zeta(c_{\mathrm{TX}}+\phi S_M)\leq\epsilon,
\end{align*}
where, from Corollary~\ref{corol1}, $\mathbb E[\left.R_k\right|\zeta]=B\zeta e^{-\zeta}$.

The DP policies are obtained after $T_{DP}{=}100$ DP iterations, and are evaluated in both \emph{Markov-}$\gamma$
and \emph{best-}$\gamma$ scenarios,
using Monte-Carlo simulation over $T{=}10^5$ slots.
The above policies in the \emph{Markov-}$\gamma$ scenario are defined similarly to
 SCDP (Sec. \ref{MarkovCOORD}) and SDDP (Sec. \ref{MarkovDIST}).
Note that COORD-SNR and DEC-SNR are non-adaptive.
On the other hand,  COORD-DP and DEC-DP adapt to the quality state $V_k$ fed back by the~FC.


 \begin{figure}[t]
\centering
\includegraphics[width = .85\linewidth,trim = 10mm 4mm 10mm 9mm,clip=true]{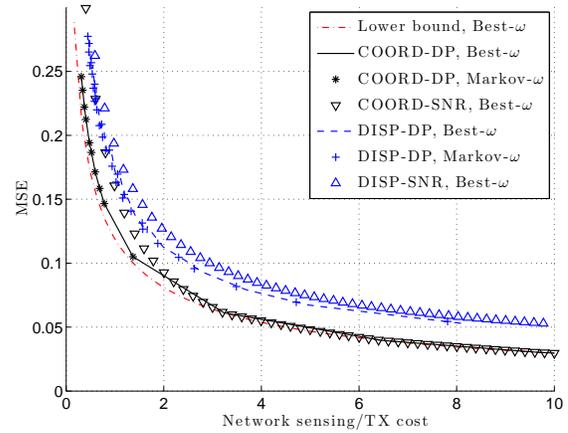}
\vspace{-3mm}
\caption{MSE as a function of the network cost, $N_S=100$.}
\label{NS10B1}
\vspace{-5mm}
\end{figure}

In Figs. \ref{NS100B1} and  \ref{NS10B1}, we plot the
 MSE (\ref{Cest}) as a function of the network cost  (\ref{CSN})
 for $N_S{=}20$ and $N_S{=}100$, respectively,  obtained by varying the parameters  $\epsilon$ and $\lambda$.
We plot also the lower bound for COORD-DP, given by Theorem~\ref{lowbound} for the \emph{best-}$\gamma$ scenario,
which is computed in the proof of Theorem~\ref{thm1} in App.~\ref{Proofthm1}.
We notice that, in the large WSN scenario, both COORD-DP and DEC-DP in the \emph{Markov-}$\gamma$ scenario (SCDP and SDDP, respectively) approach
the lower bound given by the \emph{best-}$\gamma$ scenario.
Therefore, SCDP and SDDP
perform well at a fraction of the complexity with respect to the globally optimal policy derived via DP in the \emph{Markov-}$\gamma$ scenario.
This is a result of \emph{sensing diversity}, \emph{i.e.}, in each slot, a sufficiently large pool of SNs
can sense the underlying process with high accuracy, despite the fluctuations in the \emph{local} accuracy state of each SN.
On the other hand,
DEC-DP
in the \emph{Markov-}$\gamma$ scenario (SDDP)
 incurs a small degradation in the small network scenario  with respect to the \emph{best-}$\gamma$ scenario, since, in this case,
also the SNs with lower accuracy state activate.

Interestingly, COORD-DP yields good performance also in the small network scenario.
In fact, despite the fluctuations in the accuracy state of each SN, COORD-DP always activates the
best SNs, whereas the selection is randomized and decentralized for DEC-DP.
Moreover, COORD-DP closely approaches the lower bound given by Theorem~\ref{lowbound},
and, in some cases, achieves the bound (see Theorem~\ref{thm1}). In contrast, we have verified that the lower bound of Theorem~\ref{lowbound}
for the decentralized scheme (not plotted in the figure) is loose.
This is because the lower bound of Theorem~\ref{lowbound} can be achieved only if the FC collects \emph{deterministically} a constant aggregate SNR sequence,
as dictated by the MAX-SNR scheme:
such  constant SNR sequence can be closely replicated in
 the coordinated scheme, by scheduling individually each SN and avoiding collisions; on the other hand, in the decentralized scheme, the activation decisions of the SNs are randomized and collisions occur, so that the FC experiences wide random fluctuations of
the aggregate SNR sequence around its mean.
Finally, we note that, by adapting the sensing-transmission strategy to the quality state $V_k$,
 COORD-DP and DEC-DP can achieve
 significant cost-savings
 with respect to the respective non-adaptive schemes COORD-SNR and DEC-SNR,
 up to 
74\% (for $N_S=20$) and 20\% (for $N_S=100$) for the decentralized scheme,
and up to 35\% for the coordinated one. Therefore, the maximization of the average aggregate SNR collected at the FC,
initially proposed in Sec. \ref{probform},
 is not a good design criterion,
since it does not effectively cope with the fluctuations and the stochastic dynamics
 induced by cross-layer factors such as the time-varying accuracy states, the decentralized sensing-transmission decisions of the SNs, and the channel collisions.

\begin{figure}[t]
\centering
\includegraphics[width = .85\linewidth,trim = 10mm 4mm 10mm 9mm,clip=true]{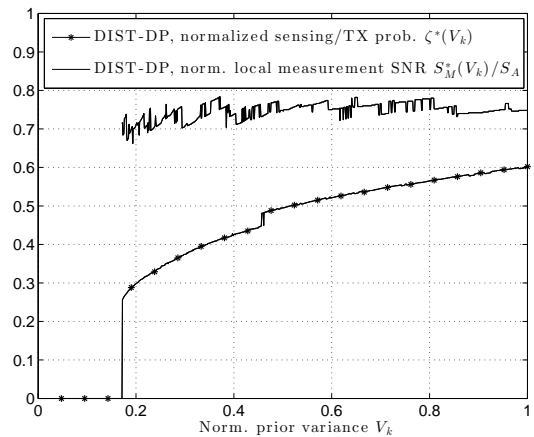}
\vspace{-3mm}
\caption{Structure of DEC-DP as a function of the prior variance $V_k$.
The corresponding simulated network cost is $1.6619$ and the MSE is $0.124$.}
\label{DISTDPSTRUC_B5}
\vspace{-5mm}
\end{figure}

In Fig.~\ref{DISTDPSTRUC_B5}, we plot the structure of DEC-DP as a function of $V_k$.
We note that, as $V_k$ increases,
\emph{i.e.}, the estimate of $X_k$ is less accurate,
 both $\zeta^*(V_k)$ and $S_M^*(V_k)$ increase,
 in order to improve the estimation accuracy ($S_M^*(V_k)$ exhibits fluctuations due to the numerical optimization).
 On the other hand, when the estimation accuracy is good ($V_k<0.2$)
  the activation probability is zero, so that the SNs can
 save energy.
 This result is in line with the myopic policy, studied in Part~II.
 Finally, note that $\zeta^*(V_k)<1,\ \forall V_k$ (Prop.~\ref{zetaopt}).

 Finally, we compare our proposed decentralized  technique to a technique proposed in \cite{Msechu}.
Therein, the estimation of a random \emph{static} parameter is considered, and decentralized censoring is employed to minimize the transmission cost of the SNs,
based on the informativeness of the measurements collected, similar to \cite{Appadwedula} for a static detection problem.
Note that (i) \cite{Msechu} assumes error-free transmissions; (ii)
it does not model the sensing cost and the ability of the SNs to tune the local measurement SNR $S_{M,n,k}$, \emph{e.g.},
by controlling the number of samples collected;
(iii) it assumes a static scenario, \emph{i.e.}, a single slot is considered and the parameter to be estimated does not vary over time.
In our framework, in contrast, (i) transmissions are prone to collisions; (ii) $S_{M,n,k}$ is a control parameter, with cost $\phi S_{M,n,k}$;
(iii) the process to be tracked is time-correlated,  and the SNs have an internal accuracy state evolving as a Markov chain.
Our proposed feedback loop enables adaptation of the sensing-transmission strategy in order to cope with the dynamics induced by these
cross-layer factors.

Since \cite{Msechu} does not consider our model exactly, we have extended it to accommodate our cross-layer dynamic setting as follows.
We denote this scheme as \emph{modified}-[17] (Mod-[17]).
Given the prior variance $V_k$ and mean $\sqrt{\alpha}\hat X_{k-1}$ of $X_k$ at the beginning of slot $k$, 
all SNs perform a measurement with common measurement SNR $S_M$. Then,
SN $n$ censors its measurement (denoted as $C_{n,k}{=}1$) if 
\begin{align}
\label{censrule}
\left|Y_{n,k}-\gamma_{n,k}\sqrt{\alpha}\hat X_{k-1}\right|<\tau\sqrt{\gamma_{n,k}^2V_k+S_A^{-1}+S_M^{-1}},
\end{align}
where the term within the square root is the variance of $Y_{n,k}$, given $(\sqrt{\alpha}\hat X_{k-1},V_k,\gamma_{n,k})$,
and transmits it otherwise ($C_{n,k}=0$).
In other words, $Y_{n,k}$ is transmitted if and only if it significantly deviates from its expected value $\sqrt{\alpha}\hat X_{k-1}$~\cite{Msechu}.
The threshold $\tau$, common to all SNs, determines the transmission probability $q$ of the SNs. From the censoring rule
(\ref{censrule}), $q=2(1-Q(\tau))$,
where $Q(x)$ is the normal Gaussian cumulative distribution function.
Note that, in this scheme, all SNs sense in each slot, so that a fixed sensing cost $\phi S_M$ is incurred, as opposed to our scheme,
where each SN either activates by sensing and transmitting or remains idle. On the other hand, transmissions occur with probability 
$q=2(1-Q(\tau))$, so that, on average, the sensing-transmission cost is $qc_{\mathrm{TX}}+\phi S_M$ in each slot.
We define the pair $(q,S_M)$ so as to optimize the aggregate SNR collected at the FC, under the sensing-transmission cost constraint,
\emph{i.e.}, using the approximation in Corollary~\ref{corol1} for the channel successes
and (\ref{Stot}), and assuming the \emph{best-}$\gamma$  scenario, 
\begin{align*}
(q^*\!\!,\!S_M^*)\!=\!\!\!\!&\underset{q\in[0,1],S_M\geq 0}{\arg\max}\!\!\!\!qN_Se^{-\frac{qN_S}{B}}\!\!\frac{S_AS_M}{S_A\!+\!S_M},
\text{s.t. }qc_{\mathrm{TX}}\!+\!\phi S_M\!\!\leq\!\!\frac{\epsilon}{N_S}\!.
\end{align*}
\begin{figure}[t]
\centering
\includegraphics[width = .85\linewidth,trim = 10mm 4mm 10mm 9mm,clip=true]{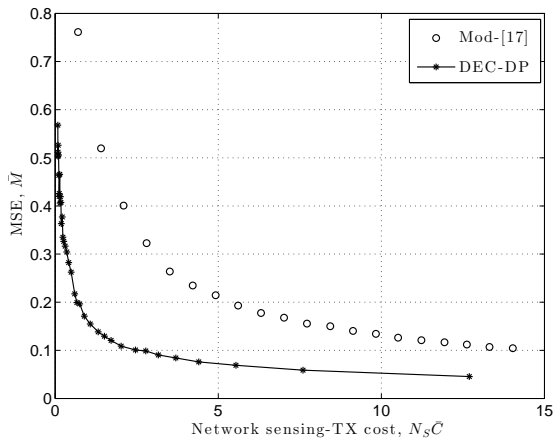}
\vspace{-3mm}
\caption{MSE as a function of the network cost, comparison between Mod-[17] and DEC-DP;
$N_S=100$, \emph{Markov-}$\gamma$ scenario.}
\label{MSE_comp}
\vspace{-5mm}
\end{figure}

Unfortunately for this scheme, the optimal estimator is not the linear Kalman filter. In fact,
censored measurements provide indirect feedback to the FC, which can be exploited to infer $X_k$.
The optimal approach is then for the FC to compute a posterior belief of $X_k$,
involving cumbersome numerical integration,
 given the measurements collected and the indirect feedback signal,
based on which an MMSE estimate of $X_k$ can be obtained.
However, note that, in our setting, the FC cannot differentiate between a censored measurement (which provides the indirect feedback signal 
$C_{n,k}{=}1$ given by (\ref{censrule})) or a collision (uncensored but lost, thus providing the indirect feedback signal $C_{n,k}{=}0$),
so that the computation of the posterior belief requires a cumbersome marginalization over these events, and over the value of 
the accuracy state $\gamma_{n,k}$ in (\ref{censrule}).
In order to overcome this difficulty, we use the idealized assumption that the FC is genie-aided, \emph{i.e.}, it knows which SN censored its measurement, as well as the accuracy state $\gamma_{n,k}$ of each SN. \emph{This information is not available to the decentralized scheme proposed in this paper, thus yielding a lower-bound to the cost-MSE trade-off achievable by Mod-[17]}. The  posterior distribution  of $X_k$, given the observations collected at the FC, the collision outcome,
 the censoring outcome and accuracy state of each SN, is evaluated numerically. Based on it, 
 the MMSE estimate of $X_k$ (posterior mean) and its posterior variance $\hat V_k$ are computed.
 Finally,
the Gaussian approximation is used, so that
the next prior belief is
 $X_{k+1}{\sim}\mathcal N(\sqrt{\alpha}\hat X_{k},\nu(\hat V_{k}))$. This scheme is then repeated in each slot.
 
\begin{figure}[t]
\centering
\includegraphics[width = .85\linewidth,trim = 10mm 4mm 10mm 9mm,clip=true]{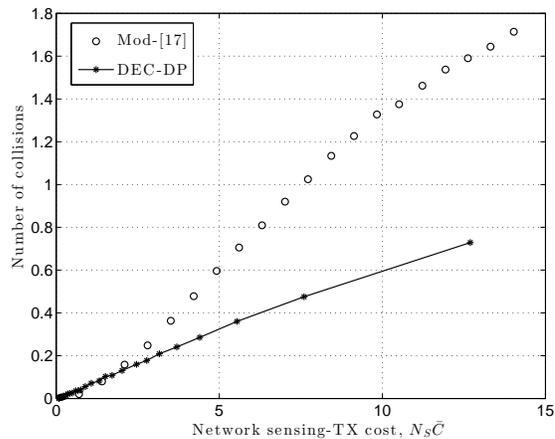}
\vspace{-3mm}
\caption{Average number of collisions per slot as a function of the network cost, comparison between Mod-[17] and DEC-DP;
 $N_S=100$.}
\label{num_coll}
\vspace{-5mm}
\end{figure}

In Fig. \ref{MSE_comp}, we evaluate the trade-off between network cost and MSE
under Mod-[17] and DEC-DP, via Monte-Carlo simulation over 3000 slots.
We notice that Mod-[17] incurs a significant performance degradation with respect to DEC-DP, despite the idealized assumption that 
the censoring and collision outcomes, as well the accuracy state of each SN, are known to the FC under Mod-[17] (such information is not available to DEC-DP).
In fact, Mod-[17] does not employ a cross-layer perspective, \emph{i.e.}, it neglects the cost of sensing (each SN senses in each slot), and the shared wireless channel,
which results in collisions and uncertainty in the number of measurements collected at the FC.
This is also confirmed by the more frequent collisions incurred by Mod-[17] with respect to DEC-DP, as shown numerically in Fig.~\ref{num_coll}.
Additionally, Mod-[17] is not designed  to cope with the time-correlated dynamics considered in our model.

 \vspace{-3mm}
\section{Conclusions}\label{conclusions}
 In this paper, we have proposed a cross-layer distributed sensing-estimation framework for WSNs,
 which exploits the quality feedback information from the FC.
 Our cross-layer design approach allows one to model the time-varying capability 
 of the SNs to accurately sense the underlying process,
  the scarce channel access resources shared by the SNs, as well as sensing-transmission costs.
We have proposed a coordinated scheme, where the FC schedules the action of each SN,
 and a more scalable decentralized scheme, where each SN performs a local decision to sense-transmit or remain idle.
 Despite the curse of dimensionality typical of the design of WSNs and multi-agent systems in asymmetric environments,
 we have
 exploited the statistical symmetry of the network and a large WSN approximation to derive structural properties of the optimal policy,
which enable a more efficient optimization via DP.
We have shown that a dense WSN provides \emph{sensing diversity}, \emph{i.e.},
only a few SNs suffice to sense accurately and transmit, with no degradation in the MSE,
 despite the local fluctuations in the observation quality.
Our analysis and numerical results show that the proposed schemes achieve near-optimal performance
also for small-medium sized WSNs, and outperform non-adaptive schemes that do not exploit the quality feedback from the FC
and a technique proposed in the literature. 
We have evaluated the communication overhead of both schemes,
proving that the decentralized one
 meets both goals of energy efficiency and scalability, requiring no coordination and minimal feedback information.

\appendices
\vspace{-3mm}
 \section{}
 \label{Proofoptsymm}
 \noindent\emph{Proof of Prop.~\ref{optsymm}:}
We refer to the optimization problem (\ref{optprob}) only. In fact, for any $\lambda>0$,
there exists $\epsilon>0$ such that the optimal policy for the problem (\ref{optproblag}) is also optimal for the problem (\ref{optprob}).
Let $\delta$ be an optimal instruction policy for (\ref{optprob}).
Let $\hat\delta$ be a new policy defined as, $\forall \mathbf D^{(O)}$,
\begin{align*}
&\hat \delta_k(\mathbf D|V_{k})\!=\!\frac{1}{|\mathcal D(\mathbf D^{(O)})|}\!\sum_{\tilde{\mathbf D}\in \mathcal D(\mathbf D^{(O)})}\!\!\!\delta_k(\tilde{\mathbf D}|V_{k}),\ 
\forall \mathbf D\in \mathcal D(\mathbf D^{(O)}).
\end{align*}
$\hat\delta$ obeys the statement of the proposition.
The distribution of the aggregate SNR collected at the FC
under the two instruction policies $\delta$ and $\hat \delta$ is identical, since the SNs are symmetric.
By induction on $k$, it follows that $\hat V_k$ has the same distribution under the two instruction policies
$\delta$ and $\hat\delta$, hence
$\bar M_{\delta}^{T}(V_0)=\bar M_{\hat{\delta}}^{T}(V_0)$.
Similarly, 
$\bar C_{\hat{\delta}}^{T,n}(V_0)=\frac{1}{N_S}\sum_{n=1}^{N_S}\bar C_{\delta}^{T,n}(V_0)\leq\frac{\epsilon}{N_S},\ \forall n$,
hence $\hat\delta$ is also optimal.
\hfill\QED

\noindent\emph{Proof of Prop.~\ref{lem2}:}
Consider two ordered instructions $\mathbf D^{(O)}=(d_{1}^{(O)},d_{2}^{(O)},\dots,d_{N_S}^{(O)})$, $\tilde{\mathbf D}^{(O)}{=}(\tilde d_{1}^{(O)},\tilde d_{2}^{(O)},\dots,\tilde d_{N_S}^{(O)}){\in}\mathcal D^{(O)}$, such that,
$d_{n}^{(O)}{=}\tilde d_{n}^{(O)}{=}(0,0,0),\forall n>t$,
 $d_{n}^{(O)}=(1,S_{M,n},B_n)$ and $\tilde d_{n}^{(O)}=(1,\tilde S_{M},B_n),\forall n\leq t$, where
 \begin{align}
\label{sm}
\tilde S_{M}=
\frac{
\sum_{m=1}^{t}A_mS_{M,m}/(S_A+S_{M,m})
}
{
\sum_{m=1}^{t}A_m/(S_A+S_{M,m})
},
\end{align}
for some $t\in\{1,2,\dots,B\}$.
If ${\mathbf D}^{(O)}$ (respectively, $\tilde{\mathbf D}^{(O)}$) is chosen in slot $k$,
then the actions $d_{n}^{(O)}$ (resp., $\tilde d_{n}^{(O)}$) are scheduled randomly to the SNs,
so that $d_n^{(O)}$ (resp., $\tilde d_{n}^{(O)}$) is assigned to SN $m$ with marginal probability $1/N_S$.
 Then,
 the aggregate SNR collected at the FC under both ${\mathbf D}^{(O)}$ and $\tilde{\mathbf D}^{(O)}$ is $\Lambda_k=\sum_{n=1}^{t}\frac{S_AS_{M,n}}{S_A+S_{M,n}}=\sum_{n=1}^{t}\frac{S_A\tilde S_{M}}{S_A+\tilde S_{M}}$.
Therefore, ${\mathbf D}^{(O)}$ and $\tilde{\mathbf D}^{(O)}$ attain the same MSE performance in slot $k$, $\hat V_k=\hat \nu(V_{k},\Lambda_k)$.
On the other hand,
 the cost for each SN under ${\mathbf D}^{(O)}$ and $\tilde{\mathbf D}^{(O)}$ satisfies
 \begin{align}
&
\mathbb E[c_{SN}(A_{n,k},S_{M,n,k})|\tilde{\mathbf D}^{(O)}]
=
\frac{t}{N_S}c_{SN}(1,\tilde S_{M})
\\&\nonumber
\leq
\frac{t}{N_S}c_{SN}\left(1,\frac{1}{t}\sum_{n=1}^{t}S_{M,n}\right)
=\mathbb E[c_{SN}(A_{n,k},S_{M,n,k})|{\mathbf D}^{(O)}],
\end{align}
where we have used the fact that (\ref{sm}) is an increasing function of $S_A$, hence $\tilde S_{M}\leq\frac{1}{t}\sum_{n=1}^{N_S}A_nS_{M,n}$,
and $c_{SN}\left(1,S_M\right)$ is increasing in $S_M$.
We conclude that a lower cost is incurred by the ordered
instruction $\tilde{\mathbf D}^{(O)}$, while achieving the same MSE accuracy as $\mathbf D^{(O)}$. The proposition is thus proved.
\hfill\QED

\noindent\emph{Proof of Prop.~\ref{lem3}:}
The target aggregate SNR $\Lambda_k$
 can be collected at the FC by scheduling $t_k>\Lambda_k/S_A$ SNs to
sense with local measurement SNR $S_M=\frac{S_A\Lambda_k}{t_kS_A-\Lambda_k}$ and to transmit.
The MSE and the next state $V_{k+1}$ is a function of the current state $V_k$ and aggregate SNR $\Lambda_k$.
Hence, given $\Lambda_k$, $t_k$ can be uniquely chosen to minimize the expected cost in slot $k$,
 $t^*(\Lambda)=\arg\min_{t}
 \frac{t}{N_S}c_{SN}\left(1,\frac{S_A\Lambda}{tS_A-\Lambda}\right)$.
Its solution yields Prop.~\ref{lem3}, but is omitted due to space constraints.
\hfill\QED

\vspace{-3mm}
\section{}
\label{Proofthm1}
\noindent\emph{Proof of Theorem~\ref{thm1}:}
From Theorem~\ref{lowbound}, $\bar M_{\delta^*}\!\!\geq\!\!\hat\nu^*(\bar \Lambda^*)$,
where
\begin{align}
\label{opwreter}
\!\!\!\bar\Lambda^*\!\!\!=\!\max_{p,S_M}\!\!\sum_{t=1}^B\!\!\frac{p(t)tS_AS_M\!(t)}{S_A+S_M(t)},\!
\text{s.t.}\!\!\sum_{t=1}^B\!\!p(t)t\!\left(c_{\mathrm{TX}}\!+\!\phi S_{M}\!(t)\right)\!\leq\!\epsilon.\!\!\!
\end{align}
Using the Lagrangian method to optimize over $S_M(\cdot)$, we have
\begin{align*}
&S_M^*(\cdot)\!=\!\arg\max_{S_M}\!\!\sum_{t=1}^Bp(t)t\!\left[\frac{S_AS_M(t)}{S_A+S_M(t)}-\mu\left(c_{\mathrm{TX}}+\phi S_{M}(t)\right)\right]\!.\!\!
\end{align*}
yielding $S_M^*(t){=}S_A\left(\frac{1}{\sqrt{\mu \phi}}-1\right)^+{\triangleq}\bar S_M,\forall t$.
Optimizing with respect to $\bar S_M$ in (\ref{opwreter}), we obtain $\bar S_{M}^*=\frac{1}{\theta}\left(\frac{\epsilon}{\bar m c_{\mathrm{TX}} }-1\right)$, where $\bar m\triangleq\sum_{t=1}^Bp(t)t$. Finally,
 optimizing over $\bar m$,
\begin{align}
&\bar\Lambda^*\!=\max_{\bar m\in [0,\min\{\epsilon/c_{\mathrm{TX}},B\}]}\frac{\bar mS_A(\epsilon-c_{\mathrm{TX}}\bar m)
}{\bar m\left(\phi S_A-c_{\mathrm{TX}}\right)+\epsilon
}.
\end{align}
Computing the derivative with respect to $\bar m$, it can be shown that the argument of the optimization is increasing in $\bar m$ if and only if
$\bar m\leq\frac{\epsilon}{c_{\mathrm{TX}}+\sqrt{\phi S_Ac_{\mathrm{TX}}}}$,
so that $\bar m^*=\min\{\frac{\epsilon}{c_{\mathrm{TX}}+\sqrt{\phi S_Ac_{\mathrm{TX}}}},B\}$.
Then,
if $\epsilon=tc_{\mathrm{TX}}(1+\sqrt{\theta S_A})$, for some $t=1,2,\dots,B$, as in the statement of Theorem~\ref{thm1},
we obtain $\bar m^*=t$, $\bar S_{M}^*=\sqrt{S_A/\theta}$, hence
 $\bar\Lambda^*=t\frac{S_AS_M^*}{S_A+S_M^*}$.
 If $\epsilon>Bc_{\mathrm{TX}}(1+\sqrt{\theta S_A})$,
 we obtain $\bar m^*=B$, $\bar S_{M}^*=\frac{1}{\theta}\left(\frac{\epsilon}{Bc_{\mathrm{TX}}}-1\right)$, hence
 $\bar\Lambda^*=B\frac{S_AS_M^*}{S_A+S_M^*}$.
The achievability of the lower bound follows from the following Prop.~\ref{lem1}, when $\Lambda_k=\bar\Lambda^*,\ \forall k$.
\begin{propos}
\label{lem1}
Let $V_0{=}1$ and $\Lambda_0^{T}{=}\bar \Lambda\mathbf 1_{T+1}$ be a constant sequence,
where $\mathbf 1_m$ is the $m$-dimensional vector of ones. Then,  
\begin{align}
\lim_{T\to\infty}R_T(1;\bar \Lambda\mathbf 1_{T+1})=\hat\nu^*(\bar \Lambda).
\end{align}
\end{propos}
\noindent\emph{Proof:}
Due to space constraints, a proof outline is provided. 
Note that $\hat\nu^*(\bar \Lambda)$ is a fixed point of 
$\hat V_k{=}\hat \nu(\nu(\hat V_{k-1}),\bar\Lambda){=}\hat V_{k-1}$, so that, if $\hat V_{k-1}{=}\hat\nu^*(\bar \Lambda)$ and $\Lambda_k{=}\bar\Lambda$, then $\hat V_k{=}\hat V_{k-1}{=}\hat\nu^*(\bar \Lambda)$.
First, we show by induction that $\{\hat V_k,k{\geq}0\}$ is a strictly decreasing sequence and $\hat V_k{>}\hat\nu^*(\bar \Lambda),\forall k$. 
In fact, let $\hat V_k{\in}(\hat\nu^*(\bar \Lambda),1]$ (this is true for $k{=}0$, since $V_0{=}1$). 
Since $\hat\nu^*(\Lambda)$ is a decreasing function of $\Lambda$,
there exists a unique $\hat\Lambda{\in}(0,\bar \Lambda)$ such that  $\hat V_k{=}\hat\nu^*(\hat\Lambda){=}\hat \nu(\nu(\hat V_{k}),\hat\Lambda)$,
hence
\begin{align}
\hat V_{k+1}=\hat\nu(\nu(\hat V_{k}),\bar\Lambda)
<\hat\nu(\nu(\hat V_{k}),\hat\Lambda)=\hat V_{k},
\end{align}
since $\hat\nu(V,\Lambda)$ is a decreasing function of $\Lambda$.
Since $\hat\nu(\nu(\hat V),\Lambda)$ is increasing in $\hat V$ and $\hat V_k>\hat\nu^*(\bar\Lambda)$, we obtain
\begin{align}
\hat\nu^*(\bar\Lambda)=\hat\nu(\nu(\hat\nu^*(\bar\Lambda)),\bar\Lambda)<
\hat\nu(\nu(\hat V_k),\bar\Lambda)=\hat V_{k+1},
\end{align}
hence $\hat V_k{\in}(\hat\nu^*(\bar \Lambda),\hat V_{k+1})$.
It follows that
 $\lim_{k\to\infty}\hat V_k{=}\hat\nu^*(\bar \Lambda)$ and $\lim_{T\to\infty}R_T(1;\bar \Lambda\mathbf 1_{T+1}){=}\hat\nu^*(\bar \Lambda)$.
\hfill\QED

\noindent Theorem~\ref{thm1} is thus proved.
\hfill\QED

\vspace{-3mm}
 \section{}
 \label{Proofthm2}
\noindent\emph{Proof of Theorem~\ref{thm2}:}
The equality $\bar C_{\delta^*}^{\gamma_{\max}}{=}\bar C_{\delta^*}$ is trivial, since
the sequence $\{t_k^*,S_{M,k}^*,k{\geq}0\}$ is common to the \emph{best-}$\gamma$ and \emph{Markov-}$\gamma$ scenarios.
Let $\Lambda_0^T$ and $\Lambda_0^{T,*}$ be the realization of the aggregate SNR sequence collected at the FC in the \emph{Markov-}$\gamma$ and \emph{best-}$\gamma$ scenarios, respectively, when
 $\{t_k^*,S_{M,k}^*,k{\geq}0\}$ is scheduled.
Let $Q_k{=}\chi(\text{less than $B$ SNs have accuracy $\gamma_{n,k}{=}\gamma_{\max}$})$.
Then,  $(1{-}Q_k)\Lambda_k^*{\leq}\Lambda_k{\leq}\Lambda_k^*,\forall k$.
In fact, if $Q_k{=}0$, then at least $B$ SNs have the best accuracy state,
and the FC will schedule $t_k^*{\leq}B$ of those SNs to activate, so that
 $\Lambda_k{=}\Lambda_k^*$.
Let $V_k$, $\hat V_k$ and $V_k^*$, $\hat V_k^*$ be the prior and posterior variances in slot $k$
in the \emph{Markov-}$\gamma$  and \emph{best-}$\gamma$ scenarios, respectively, so that 
$\bar M_{\delta^*}^{T}=\frac{1}{T+1}\mathbb E\left[\sum_{k=0}^T\hat V_k|V_0\right]$ and
$\bar M_{\delta^*}^{T,\gamma_{\max}}=\frac{1}{T+1}\sum_{k=0}^T\hat V_k^*$.
Note that, since $\Lambda_k{\leq}\Lambda_k^*,\forall k$, then $\hat V_k{\geq}\hat V_k^*,\forall k$, from which
the left-hand inequality in Theorem \ref{thm2} follows.
Let $\bar k$ be a slot index such that $Q_{\bar k-1}{=}1$ and $Q_j{=}0,\forall j{=}\bar k,\bar k{+}1,\dots,\bar k{+}J{-}1$, for some $J{>}0$.
Since $\Lambda_k{=}\Lambda_k^*$ when $Q_k{=}0$, we have
\begin{align}
\sum_{k=\bar k}^{\bar k+J-1}\hat V_k=
\sum_{k=\bar k}^{\bar k+J-1}\hat\nu^{k-\bar k}(V_{\bar k};\Lambda_{\bar k}^{k,*}).
\end{align}
Since $\hat\nu^{k-\bar k}(V_{\bar k};\Lambda_{\bar k}^{k,*})$ is an increasing concave function of $V_{\bar k}$,
and $1\geq V_{\bar k}\geq V_{\bar k}^*\geq 1-\alpha$,
 using (\ref{derivV}) and $D_n\geq 1$ we obtain
\begin{align*}
&\sum_{k=\bar k}^{\bar k+J-1}\hat V_k\leq\!\!\!\!
\sum_{k=\bar k}^{\bar k+J-1}\left[\hat V_k^*
+
\left.\frac{\mathrm d\hat\nu^{k-\bar k}(v;\Lambda_{\bar k}^{k,*})}{\mathrm d v}\right|_{v=V_{\bar k}^*}\!\!\!\!\!\!\!\!\!\!(V_{\bar k}-V_{\bar k}^*)\right]
\nonumber\\&
\leq
\sum_{k=\bar k}^{\bar k+J-1}\left[\hat V_k^*+\alpha^{k-\bar k+1}\right]
\leq
\hat V_{\bar k}^*+\frac{\alpha}{1-\alpha}+\sum_{k=\bar k+1}^{\bar k+J-1}\hat V_k^*.
\end{align*}
By using the inequality $\hat V_k\leq 1$ when $Q_k=1$, we obtain
\begin{align*}
&\sum_{k=0}^{T}\hat V_k\leq
(1-Q_0)\hat V_0^*
\!+\!\sum_{k=1}^{T}\!\left[
(1-Q_k)(1-Q_{k-1})\hat V_k^*+Q_k
\right]
\nonumber\\&
\!+\!Q_0\!+\!\sum_{k=1}^{T}\!(1\!-\!Q_k)Q_{k-1}\!\left(\hat V_{k}^*\!+\!\frac{\alpha}{1-\alpha}\right)
\!\!=
\!\!\sum_{k=0}^{T}\hat V_k^*
\!+Q_0(1\!-\!\hat V_0^*)
\nonumber\\&
\!\!+\!\!\sum_{k=1}^{T}\!
\left[
\frac{\alpha(1-Q_k)Q_{k-1}}{1-\alpha}
+Q_k(1-\hat V_k^*)
\right]
\!\leq\!
  \sum_{k=0}^{T}\!\left(\hat V_k^*+\frac{Q_k}{1-\alpha}\right)\!.
\end{align*}
Assuming
$\boldsymbol{\gamma}_k$ is at steady-state,
and letting
\begin{align*}
\mathcal Q\triangleq \mathbb P(Q_k=1)\!
=\!\sum_{i=0}^{B-1}\left(\begin{array}{c}\!\!\!N_S\!\!\!\\i\end{array}\right)\pi_{\gamma}(\gamma_{\max})^i
(1\!-\!\pi_{\gamma}(\gamma_{\max}))^{N_S-i},
\end{align*}
we obtain
$\bar M_{\delta^*}^{T}=\mathbb E\!\left[R_T(V_0;\Lambda_0^T)|V_0\right]\leq
\bar M_{\delta^*}^{T,\gamma_{\max}}\!+\!\frac{\mathcal Q}{1\!-\!\alpha}
$. Finally, 
(\ref{inex}) follows from
$\mathcal Q\leq \exp\left\{-\frac{\left(N_S\pi_{\gamma}(\gamma_{\max})-B+1\right)^2}{2N_S\pi_{\gamma}(\gamma_{\max})}\right\}$ (Chernoff's inequality)
 when $N_S\pi_{\gamma}(\gamma_{\max})\geq B-1$.
\hfill\QED

\vspace{-3mm}
\section{}
\label{proofoflemchannel}
\noindent\emph{Proof of Prop.~\ref{psr}:}
 Let $U(t,b)$ be the number of combinations of $t$ transmissions over $b$ channels, all unsuccessful.
 We have $U(t,1){=}1{-}\chi(t{=}1)$, since 
 the transmission is successful if and only if $t{=}1$, when $b{=}1$.
For $b>1$, we have the recursion
\begin{align*}
&U(t,b)=\sum_{n=0,n\neq 1}^t\left(\begin{array}{c}t\\n\end{array}\right)U(t-n,b-1),
\end{align*}
\emph{i.e.}, $n$ SNs transmit in the first channel (where $n\neq 1$, otherwise 
a successful transmission occurs), and the remaining $t-n$ SNs in the remaining $b-1$ channels.
By induction, it can be proved that, for $t\geq 0$, $b\geq 1$,
\begin{align}
\nn
&U(t,b)\!=\!\!\!\!\!\!\!\!\!\!\sum_{k=0}^{\min\{t,b-1\}}\!\!\!\!\!\!\!\!(-1)^k\!\!
\left(\begin{array}{c}\!\!b\!\!\\\!\!k\!\!\end{array}\right)\!\!
\frac{t!}{(t-k)!}
(b-k)^{t-k}
\!+b!\chi(t=b)(-1)^b,
\\&
\text{hence }p_{R|T}(r|t)=\left(\begin{array}{c}B\\r\end{array}\right)\frac{t!}{(t-r)!}\frac{U(t-r,B-r)}{B^t},
\end{align}
since there are $B!/r!/(B-r)!$ combinations 
of $r$ channels where the transmission is successful (\emph{i.e.}, one and only one SN transmits),
$\frac{t!}{(t-r)!}$ ways of selecting $r$ SNs to transmit in the successful channels, and 
$U(t-r,B-r)$ combinations of allocating the $t-r$ remaining nodes to the $B-r$ unsuccessful channels;
$B^t$ is the number of combinations to allocate $t$ SNs to $B$ channels.
$p_R(r;q)$ is then given by
\begin{align*}
p_R(r;q)=
\sum_{t=r}^{N_S}
\left(\begin{array}{c}\!\!\!\!N_S\!\!\!\!\\\!\!\!\!t\!\!\!\!\end{array}\right)
q^t(1-q)^{N_S-t}
\left(\begin{array}{c}\!\!\!B\!\!\!\\\!\!\!r\!\!\!\end{array}\right)\frac{t!U(t-r,B-r)}{(t-r)!B^t},
\end{align*}
yielding Prop.~\ref{psr} after algebraic manipulation.
\hfill\QED

\noindent\emph{Proof of Prop.~\ref{zetaopt}:}
Let $\zeta^{(1)}>1$ and $\zeta^{(2)}=1$.
We show that the cost-to-go function $\bar W^{T-k}(V_{k})$ computed under
$\zeta^{(2)}$ lower bounds the cost-to-go function computed under $\zeta^{(1)}$, for any value of the SNR $S_M$,
so that, necessarily, the minimizer of the DP stage (\ref{DPzeta}) is such that $\zeta_{k}^*(V_k)\leq 1$.
Neglecting additive and multiplicative terms independent of $\zeta$ and letting $S_T=\frac{S_AS_M}{S_A+S_M}$,
we write the cost-to-go function under a generic $\zeta$ as
\begin{align}
\label{DPzeta2}
&f(\zeta)
\triangleq
-\sum_{r=1}^B\mathbb P(R_k\geq r|\zeta)
\\&
\times\left[\hat \nu\left(V_{k},(r-1)S_T\right)-\hat \nu\left(V_{k},rS_T\right)-(W(r-1)-W(r))\right],
\nonumber
\end{align}
where $W(r){\triangleq}\bar W^{T-k-1}\left(\nu\left(\hat \nu\left(V_{k},rS_T\right)\right)\right)$.
It can be proved by induction that 
$W(r{-}1){>}W(r)$. Hence, from (\ref{DPzeta2})
we obtain $f(\zeta^{(2)}){<}f(\zeta^{(1)})$ since
 $\mathbb P(R_k\geq r|\zeta^{(2)}){\geq}\mathbb P(R_k\geq r|\zeta^{(1)}),\forall r$ (from Corollary \ref{corol1})
 and $\hat \nu\left(V_{k},(r-1)S_T\right){>}\hat \nu\left(V_{k},rS_T\right)$,
 thus proving the proposition.
\hfill\QED
\vspace{-3mm}
  \section{}
  \label{appprop}
\begin{propos}\label{propV}
$\hat\nu^T(V_0;\Lambda_0^{T})$ and $R_T(V_0;\Lambda_0^T)$ are convex functions of $\Lambda_0^T$, decreasing in $\Lambda_k$,
 concave increasing in $V_0$.
\end{propos}
\noindent\emph{Proof:}
We prove the property for $\hat\nu^k(V_0;\Lambda_0^{k})$. 
The same property holds for
$R_T(V_0;\Lambda_0^T)$, using (\ref{RT}).
Let $\mathbf X_k=[N_k, D_k]^T$ be defined recursively as 
$\mathbf X_{-1}=\left[\frac{V_0-(1-\alpha)}{\alpha}, 1\right]^T$ and, for $k\geq 0$,
$\mathbf X_{k}=\mathbf P_k\mathbf X_{k-1}$, where
\begin{align}
\mathbf P_k=\left[
\begin{array}{cc}
\alpha & 1-\alpha
\\
\alpha\Lambda_k&1+(1-\alpha)\Lambda_k
\end{array}\right].
\end{align}
Then, it can be shown by induction, by using the update equations $\nu(\cdot)$, $\hat \nu(\cdot)$ in (\ref{nu}) and (\ref{nu2}),
that $\hat V_k=N_k/ D_k$.
We have $\mathbf X_{k}=\mathbf P_{k:0}\mathbf X_{-1}$,
where $\mathbf P_{k:i}=\mathbf P_k\times\mathbf P_{k-1}\times\dots\times \mathbf P_i$, for $k\geq i$.
Notice that $\mathbf X_0{=}(V_0,1{+}V_0\Lambda_0)^T{>}0$ (non-negative entries),
so that $\mathbf X_i{>}0$ (entry-wise) by induction.
The derivative of $\hat\nu^k(V_0;\Lambda_0^{k})$ with respect to $\Lambda_i$ is given by
\begin{align}
\label{derivSi}
&\frac{\mathrm d\hat\nu^k(V_0;\Lambda_0^{k})}{\mathrm d \Lambda_i}
=-\frac{1}{ D_k^2}
\mathbf X_{k}^T
\left[
\begin{array}{cc}
0 & 1
\\
-1&0
\end{array}\right]
\frac{\mathrm d\mathbf X_{k}}{\mathrm d\Lambda_i}
\\&\nonumber
=
\!-
\mathbf X_{k}^T
\left[
\begin{array}{cc}
0 & 1
\\
-1&0
\end{array}\right]
\mathbf P_{k:i+1}
\left[\begin{array}{c}0\\1\end{array}\right]
\frac{N_{i}}{ D_k^2}
=-\alpha^{k-i}\frac{N_i^2}{ D_k^2}<0,
\end{align}
where the last equality follows by induction on $k$. 
 Therefore, $\hat\nu^k(V_0;\Lambda_0^{k})$ is a decreasing function of $\Lambda_i$.
We now compute the Hessian matrix $\mathbf H$ of $\hat\nu^k(V_0;\Lambda_0^{k})$,  with components 
$\mathbf H_{i,j}=\frac{\mathrm d^2\hat\nu^k(V_0;\Lambda_0^{k})}{\mathrm d \Lambda_i\mathrm d \Lambda_j}$.
For $j\geq i$ (the case $j<i$ is obtained by symmetry of $\mathbf H$), since $N_i$ is independent of $\Lambda_j$,
from (\ref{derivSi})
 we obtain 
\begin{align*}
 \mathbf H_{i,j}
\!\!=\!\!\frac{\!\mathrm d^2\hat\nu^k(V_0\!;\!\Lambda_0^{k})\!}{\mathrm d \Lambda_i\mathrm d \Lambda_j}
\!\!=\!\!
\frac{\!2N_i^2\alpha^{k-i}\!\!\!\!\!\!\!\!}{ D_k^3}[0,\!1]\frac{\mathrm d\mathbf X_k}{\mathrm d\Lambda_j}
\!\!=\!\!
\frac{\!2N_i^2\!N_{j}\alpha^{k-i}\!\!\!\!\!\!\!\!}{ D_k^3}
[0,\!1]\mathbf P_{k:j+1}\!\!
\left[\begin{array}{c}\!\!\!0\!\!\!\\\!\!\!1\!\!\!\end{array}\right]\!\!.
\end{align*}
Let $\mathbf D$ be a $(k+1)\times(k+1)$ diagonal matrix with diagonal entries
$\mathbf D_{i,i}=N_i^2\alpha^{k-i}$. Then,
\begin{align}
\label{f}
[\mathbf D^{-1}\mathbf H\mathbf D^{-1}]_{i,j}
=
\frac{2}{ D_k^3}
\frac{[0,1]\mathbf P_{k:j+1}[0,1]^T}
{N_j\alpha^{k-j}}\triangleq f_j.
\end{align}
%
 Note that $\mathbf D^{-1}\mathbf H\mathbf D^{-1}=
\mathbf E
\mathbf F\mathbf E^T
$, where $\mathbf E$ is an upper-triangular matrix
with all non-zero entries equal to $1$ on the diagonal and upper off-diagonal entries,  
and all other entries equal to zero, 
%
 and $\mathbf F$ is a diagonal matrix with diagonal elements $\mathbf F_{i,i}=f_i-f_{i+1},i<k$ and $\mathbf F_{k,k}=f_k$, 
%
 Finally, we obtain
$\mathbf H=
(\mathbf D\mathbf E)
\mathbf F(\mathbf D\mathbf E)^T$,
and therefore $\mathbf H$ is positive definite if and only if 
$\mathbf F$ is, that is, 
if and only if 
$f_i>f_{i+1},\forall i<k$ and $f_k>0$. 
From (\ref{f}) we have $f_k=\frac{2}{ D_k^3N_k}>0$.
On the other hand, for $i<k$,
$f_i>f_{i+1}$ is equivalent to
\begin{align}
[0,1]\mathbf P_{k:i+1}[0,1]^T
N_{i+1}
>
\alpha
[0,1]\mathbf P_{k:i+2}[0,1]^T
N_i,
\end{align}
and,
using the fact that $\mathbf P_{k:i+1}
=\mathbf P_{k:i+2}\mathbf P_{i+1}$,
$N_{i+1}=[\alpha,1-\alpha]\mathbf X_i$ and 
$N_{i}=[1,0]\mathbf X_i$,
we obtain
\begin{align}
&[0,1]\mathbf P_{k:i+2}
\left[\mathbf P_{i+1}[0,1]^T[\alpha,1-\alpha]
-[0,1]^T[\alpha,0]\right]
\mathbf X_i
\nonumber\\&
=
(1-\alpha)[0,1]\mathbf P_{k:i+2}
\mathbf P_{i+1}\mathbf X_i
=
(1-\alpha) D_k
>
0,
\end{align}
hence $\hat\nu^k(V_0;\Lambda_0^{k})$ is convex with respect to $\Lambda_0^k$.
We have
\begin{align}
\label{derivV}
&\frac{\mathrm d\hat\nu^k(V_0;\Lambda_0^{k})}{\mathrm d V_0}
=
-\frac{1}{ D_k^2}
\mathbf X_{k}^T
\left[
\begin{array}{cc}
0 & 1
\\
-1&0
\end{array}\right]
\frac{\mathrm d\mathbf X_{k}}{\mathrm dV_0}
\\\nonumber
&=
-\frac{1}{ D_k^2}
[
V_0,1+V_0\Lambda_0
]
\mathbf P_{k:1}^T\!\!
\left[
\begin{array}{cc}
0 & 1
\\
-1&0
\end{array}\right]\!\!
\mathbf P_{k:1}\!\!
\left[
\begin{array}{c}
\!\!\!1\!\!\!\\\!\!\!\Lambda_0\!\!\!
\end{array}
\right]
\!=\!
\frac{\alpha^{k}}{ D_k^2}>0,
\end{align}
where the last step follows by induction. Furthermore,
\begin{align*}
\frac{\mathrm d^2\hat\nu^k(V_0;\Lambda_0^{k})}{\mathrm d V_0^2}
\!\!=\!\!-\frac{2\alpha^{k}}{ D_k^3}[0,1]\frac{\mathrm d\mathbf X_{k}}{\mathrm dV_0}
\!=\!
-\frac{2\alpha^{k}}{D_k^3}[0,1]
\mathbf P_{k:1}\!\!
\left[
\begin{array}{c}
\!\!\!1\!\!\!\\\!\!\!\Lambda_0\!\!\!
\end{array}
\right]\!\!<\!\!0,
\end{align*}
thus proving that $\hat\nu^k(V_0;\Lambda_0^{k})$ is concave increasing in $V_0$.
\hfill\QED

\noindent The next theorem 
addresses the optimization problem
\begin{align}
\label{RTstar}
\!\!\!\!R_T^*(\bar\Lambda)=\!\!\!\min_{\Lambda_0^T,\Lambda_k\geq 0,\forall k} R_T(1;\Lambda_0^{T}),
\ \text{s.t.}\ 
\frac{1}{T+1}\sum_{k=0}^T\Lambda_k=\bar\Lambda,
\end{align}  
whose minimizer is denoted as $\Lambda_0^{T*}(\bar\Lambda)$.
  To this end, we define $\Lambda_{0,-1}^*=\Lambda_{1,-1}^*=\infty$ and, for $m\geq 0$,
 \begin{align*}
\Lambda_{0,m}^*\!\!\triangleq\!\!\frac{\sqrt{\frac{1-\alpha^{m+2}}{1-\alpha^{m+1}}}-1}{1-\alpha},\ 
\Lambda_{1,m}^*\!\!\triangleq\!\!\Lambda_{0,m}^*\left(\!\!1\!-\!\alpha\sqrt{\frac{1-\alpha^{m+1}}{1-\alpha^{m+2}}}\right)\!\!.
\end{align*}
 \begin{thm} 
 \label{thm3}
 Let $\bar \Lambda>0$, and let $m\geq 0$ uniquely solve
 \begin{align}
\label{cond}
\!\!\!\!\!\frac{\Lambda_{0,m}^*\!\!\!+\!(T\!-\!m\!-\!1)^+\!\Lambda_{1,m}^*}{T+1}\!\!\leq\!\!\bar \Lambda\!\!<\!\!\frac{\Lambda_{0,m-1}^*\!\!+\!(T\!-\!m)^+\!\Lambda_{1,m-1}^*}{T+1}.\!\!\!
\end{align}
Then, if $m\geq T$, and omitting the dependence of $\Lambda_0^{T*}$ on $\bar \Lambda$,
\begin{align}
&\Lambda_0^*=(T+1)\bar \Lambda\in [\Lambda_{0,m}^*,\Lambda_{0,m-1}^*),\ \Lambda_k^*=0,\ k>0;
\\&
\label{RT1}
R_T^*(\bar\Lambda)=1-\frac{1}{T+1}\frac{1-\alpha^{T+1}}{1-\alpha}\frac{(T+1)\bar \Lambda}{1+(T+1)\bar \Lambda}.
\end{align}
Otherwise,
\begin{align*}
\!\!R_T^*(\bar\Lambda)
\!\!=\!\!
1\!-\!\frac{T\!-\!m}{T\!+\!1}\hat v^*(\Lambda_1^*)
\!-\!\frac{1-\alpha^{m+1}}{(T\!+\!1)(1\!-\!\alpha)}
\frac{\nu(\hat v^*(\Lambda_1^*))}
{1\!+\!\nu(\hat v^*(\Lambda_1^*))\Lambda_{T-m}^*\!\!\!\!\!\!\!},
\end{align*}
where $\Lambda_{k}^*=0,\ \forall k\geq T-m+1$,
 \begin{align}
\label{Si}
&\Lambda_k^*=(1-\alpha)\Lambda_0^*\frac{1+\Lambda_0^*}{1+(1-\alpha)\Lambda_0^*},\ k\leq T-m-1,\\
&\Lambda_{T-m}^*\!\!=\!\!\frac{1+\Lambda_0^*}{1\!+\!(1\!-\!\alpha)\Lambda_0^*}\!\!\left[\!\!\sqrt{1\!-\!\alpha^{m+1}}\sqrt{(1\!+\!\Lambda_0^*)^2\!-\!\alpha\Lambda_0^{*2}}\!-\!1\!\right],
\nonumber
\end{align}
and $\Lambda_0^*{\in}[\Lambda_{0,m}^*,\Lambda_{0,m-1}^*)$ uniquely solves, for $\Lambda_1^*$ and $\Lambda_{T-m}^*$ given by (\ref{Si}),
 $\bar \Lambda{=}\frac{\Lambda_0^*+(T-m-1)\Lambda_1^*+\Lambda_{T-m}^*}{T+1}$.
  \end{thm}
\noindent\emph{Proof:}
Note that there is a one-to-one mapping between $\Lambda_0^{T*}(\bar\Lambda)$ defined in the theorem
 and $\bar \Lambda$.
 In fact, $\Lambda_k^*$ is a non-decreasing, continuous function of $\Lambda_0^*$, so that the sample mean
 $\bar \Lambda=\frac{1}{T+1}\sum_{i=0}^T \Lambda_i^*$ is an increasing function of $\Lambda_0^*$.
 Moreover, the condition (\ref{cond}) is equivalent to  $\Lambda_0^*\in[\Lambda_{0,m}^*,\Lambda_{0,m-1}^*)$.
 Therefore, we can equivalently prove that, for any $T{\geq}0,m{\geq}0,\Lambda_0^*\in[\Lambda_{0,m}^*,\Lambda_{0,m-1}^*)$,
$\Lambda_0^{T,*}$ as defined in the theorem minimizes $R_T(1;\Lambda_0^T)$ among all the SNR sequences with sample mean 
$\bar \Lambda=\frac{1}{T+1}\sum_{i=0}^T \Lambda_i^*$.
 
 Let $m\geq 0,T\geq 0,\Lambda_0\in[\Lambda_{0,m}^*,\Lambda_{0,m-1}^*)$,
 and $\Lambda_1^T$ as in the theorem.
We have 
  $\frac{\mathrm dR_T(1;\Lambda_0^T)}{\mathrm d\Lambda_i}=-Z_i^2$,
 where
\begin{align}
\label{Zi}
Z_i\triangleq \sqrt{\frac{1}{T+1}\sum_{k=i}^{T}\alpha^{k-i}\frac{N_i^2}{ D_k^2}}
\end{align}
 (see proof of Prop.~\ref{propV} in App.~\ref{appprop}).
Since $R_T(\Lambda_0^T)$ is a convex function of $\Lambda_0^T$ (Prop.~\ref{propV}) and $\Lambda_i=0,\forall i\geq (T-m)^++1$, $\Lambda_0^T$ is optimal  if and only if 
$\sum_{i=0}^{T}\beta_iZ_i^2\leq 0$,
for all $\beta_0^T$ such that $\sum_{i=0}^{T}\beta_i=0$ (due to sample mean constraint)
and $\beta_i\geq 0,\forall i\geq (T-m)^++1$.
Equivalently, using ,$\beta_0=-\sum_{i=1}^{T}\beta_i$,
\begin{align}
\beta_0Z_0^2+\sum_{i=1}^{T}\beta_iZ_i^2=
\sum_{i=1}^{T}\beta_i(Z_i^2-Z_0^2)\leq 0,
\end{align}
for all vectors $\beta_1^T$ such that $\beta_i\geq 0,\forall i\geq (T-m)^++1$, \emph{i.e.},
\begin{align}
\label{48}
\!\!\!\!Z_i=Z_{i-1},\forall i\!\leq\!(T-m)^+,\ 
Z_i\leq Z_{T-m},\forall i\!>\!(T-m)^+\!\!.\!\!
\end{align}
By rearranging the terms and using the expression of $Z_i$ in (\ref{Zi}), (\ref{48}) is equivalent to
\begin{align}
\label{ddd}
&\frac{\alpha^{i+1}\hat V_i^2}{N_{i+1}^2-\alpha N_i^2}
=
\sum_{k=i+1}^{T}\frac{\alpha^{k}}{ D_k^2},\ \forall i\leq (T-m)^+-1,
\end{align}
Equivalently, $Z_{i}\leq Z_{(T-m)^+},\ \forall i\geq T-m+1$, and
\begin{align}
\label{aaa}
&\!\!\!\frac{\hat V_{T-m-1}^2}{1-\alpha N_{T-m-1}^2/N_{T-m}^2\!\!\!\!\!\!\!\!}
=
\hat V_{T-m}^2\frac{1-\alpha^{m+1}\!\!\!\!}{1-\alpha},\text{ if }m\leq T-1,
\\
\label{bbb}
&\!\!\!\frac{\hat V_{i-1}^2}{\!1\!-\!\alpha N_{i-1}^2/N_{i}^2\!}
\!\!=\!\!
\frac{\hat V_{i}^2}{1\!-\!\alpha N_{i}^2/N_{i+1}^2\!\!\!\!},
\forall i\!\!\leq\!\!T\!-\!m\!-\!1,\text{if }\!m\!<\!T\!-\!1,\!\!\!
\end{align}
where in (\ref{aaa}) we have used the fact that $\Lambda_k{=}0,\forall k{\geq}T{-}m{+}1$, hence $ D_k{=}D_{k-1}{=}D_{T-m}$;
in (\ref{bbb}) we have combined the equations (\ref{ddd}) for $i$ and $i{+}1$.
From (\ref{bbb})
for $i{=}1,2,\dots,T{-}m{-}1$, note that $\hat V_i{>}\hat V_{i-1}$ if and only if $N_{i}^2{<}N_{i-1}N_{i+1}$. 
 This in turn is equivalent to $\hat V_i{<}\hat V_{i-1}$,
thus we must necessarily have $\hat V_i{=}\hat V_{i-1}{=}\hat V_0,\forall i{=}1,2,\dots,T{-}m{-}1$,
and therefore $\Lambda_i{=}\Lambda_1,\forall i{=}1,2,\dots,T{-}m{-}1$ and
$\hat V_i{=}\hat\nu^*(\Lambda_1),\forall i{=}0,1,\dots,T{-}m{-}1$.
It follows that, for a given $\Lambda_0\in[\Lambda_{0,m}^*,\Lambda_{0,m-1}^*)$ with $m<T-1$ and $V_0=1$, we have $\hat V_0=1/(1+\Lambda_0)$.
Then, (\ref{bbb}) implies
\begin{align*}
\hat V_i
=
\hat\nu(\nu(\hat V_{i-1}),\Lambda_i)
=
\hat\nu(\nu(\hat V_{0}),\Lambda_1)
=\hat V_0,
\ \forall i\leq T-m-1,
\end{align*}
yielding
$
\Lambda_i=\Lambda_1
=
\Lambda_0
(1-\alpha)\frac{1+\Lambda_0}
{
1+(1-\alpha)\Lambda_0
},\ \forall i\leq T-m-1.
$
thus proving the optimality of  (\ref{Si}) for $i\leq T-m-1$.

Finally, using (\ref{aaa}) and the fact that $\hat V_{T-m-1}=\hat V_0$, we have
\begin{align*}
&\hat V_{T-m}^2\frac{1-\alpha^{m+1}}{1-\alpha}
=
\frac{\hat V_{0}^2}{1-\alpha\Lambda_{0}^2/\Lambda_{1}^2}
=
\frac{\hat V_{0}^2}{1-\alpha}\frac{(1-\alpha(1-\hat V_0))^2}{1-\alpha(1-\hat V_0)^2},
%
\nn\\
&\Rightarrow\Lambda_{T-m}
=
 \frac{\frac{1}{\hat V_0}\sqrt{1-\alpha^{m+1}}\sqrt{1-\alpha(1-\hat V_0)^2}-1}{1-\alpha(1-\hat V_0)},
\end{align*}
yielding (\ref{Si}).
To obtain a feasible solution, we must have $\Lambda_{T-m}\geq 0$, \emph{i.e.},
$\!\Lambda_0\!\geq\Lambda_{0,m}^*,
$
which holds by assumption.

Finally, we prove by induction on $m$ that
$Z_{i}{\leq}Z_{(T-m)^+},\forall i$. 
 This trivially holds with equality for $i=0,1,\dots, (T-m)^+$, as proved in the first part of the proof.
Therefore, we need to prove the inequality for $i\geq (T-m)^++1$.
We have that $Z_i$ is a continuous function of $\Lambda_0$.
Now, let $m\geq 0$ and assume that $Z_{i}\leq Z_{(T-m)^+},\forall \Lambda_0\in[\Lambda_{m,1}^*,\Lambda_{m-1,0}^*)$.
We show that this implies that $Z_{i}\leq Z_{(T-m-1)^+},\forall \Lambda_0\in[\Lambda_{m+1,1}^*,\Lambda_{m,0}^*)$.
Let $\Lambda_0\in[\Lambda_{m+1,1}^*,\Lambda_{m,0}^*)$. 
For $i\geq (T-m-1)^+-1$, using the fact that $\Lambda_i=0,\forall i\geq T-m$, we have
$
Z_i\!=\!\sqrt{\frac{1-\alpha^{T-i+1}}{(T+1)(1-\alpha)}}\hat V_i.
$
Using $\hat V_i=1-\alpha^{i-(T-m-1)^+}(1-\hat V_{(T-m-1)^+})$, we obtain
 \begin{align*}
&Z_{(T-m-1)^+}-Z_{i}
\!\propto\!\hat V_{(T-m-1)^+}\sqrt{1-\alpha^{T-(T-m-1)^++1}}
\nonumber\\&
-\sqrt{1-\alpha^{T-i+1}}
[1-\alpha^{i-(T-m-1)^+}(1-\hat V_{(T-m-1)^+})].
\end{align*}
By inspection, using the fact that $\hat V_{(T-m-1)^+}=\hat V_0=\frac{1}{1+\Lambda_0}$,  $Z_{(T-m-1)^+}-Z_{i}$
 is a decreasing function of $\Lambda_0$, minimized by $\Lambda_0=\Lambda_{0,m}^*$.  
 Using $\left[Z_{(T-m)^+}\!-\!Z_{(T-m-1)^+}\right]_{\Lambda_0=\Lambda_{0,m}^*}\!\!\!\!\!\!\!\!\!\!\!\!\!=0$
 and the induction hypothesis, 
 we thus obtain
 \begin{align}
&Z_{(T-m-1)^+}\!\!\!-Z_{i}
\!\geq\!
\left[Z_{(T-m-1)^+}\!\!\!-\!Z_{i}
\!=\!
Z_{(T-m)^+}\!\!\!-\!Z_{i}\right]_{\Lambda_0=\Lambda_{0,m}^*}\!\!\!\!\!\!\!\geq 0,
\nonumber
\end{align}
thus proving the induction step and the theorem.
\hfill\QED

\noindent Corollary \ref{corollary} follows from Theorem~\ref{thm3} and Prop.~\ref{lem1} in App.~\ref{Proofthm1}.
\vspace{-3mm}
\begin{corol}
\label{corollary}
$R_\infty^*(\bar\Lambda)
\triangleq\lim_{T\to\infty}R_T^*(\bar\Lambda)=\hat \nu^*\left(\bar\Lambda\right),
$
achievable by the constant aggregate SNR sequence $\Lambda_k=\bar\Lambda,\ \forall k$.
\end{corol}
\vspace{-5mm}
\bibliographystyle{IEEEtranS}
\bibliography{IEEEabrv,References} 

\begin{thebibliography}{10}
\providecommand{\url}[1]{#1}
\csname url@samestyle\endcsname
\providecommand{\newblock}{\relax}
\providecommand{\bibinfo}[2]{#2}
\providecommand{\BIBentrySTDinterwordspacing}{\spaceskip=0pt\relax}
\providecommand{\BIBentryALTinterwordstretchfactor}{4}
\providecommand{\BIBentryALTinterwordspacing}{\spaceskip=\fontdimen2\font plus
\BIBentryALTinterwordstretchfactor\fontdimen3\font minus
  \fontdimen4\font\relax}
\providecommand{\BIBforeignlanguage}[2]{{%
\expandafter\ifx\csname l@#1\endcsname\relax
\typeout{** WARNING: IEEEtranS.bst: No hyphenation pattern has been}%
\typeout{** loaded for the language `#1'. Using the pattern for}%
\typeout{** the default language instead.}%
\else
\language=\csname l@#1\endcsname
\fi
#2}}
\providecommand{\BIBdecl}{\relax}
\BIBdecl

\bibitem{Appadwedula}
S.~Appadwedula, V.~Veeravalli, and D.~Jones, ``{Decentralized Detection With
  Censoring Sensors},'' \emph{IEEE Transactions on Signal Processing}, vol.~56,
  no.~4, pp. 1362--1373, 2008.

\bibitem{Bernstein}
D.~S. Bernstein, R.~Givan, N.~Immerman, and S.~Zilberstein, ``{The Complexity
  of Decentralized Control of Markov Decision Processes},'' \emph{Math. Oper.
  Res.}, vol.~27, no.~4, pp. 819--840, Nov. 2002.

\bibitem{Bertsekas2005}
D.~Bertsekas, \emph{Dynamic programming and optimal control}.\hskip 1em plus
  0.5em minus 0.4em\relax Athena Scientific, Belmont, Massachusetts, 2005.

\bibitem{Chamberland}
J.-F. Chamberland and V.~Veeravalli, ``{Decentralized detection in sensor
  networks},'' \emph{IEEE Transactions on Signal Processing}, vol.~51, no.~2,
  pp. 407--416, 2003.

\bibitem{Gaudenzi}
R.~De~Gaudenzi, O.~del Rio~Herrero, G.~Acar, and E.~Garrido~Barrabes,
  ``{Asynchronous Contention Resolution Diversity ALOHA: Making CRDSA Truly
  Asynchronous},'' \emph{IEEE Transactions on Wireless Communications},
  vol.~13, no.~11, pp. 6193--6206, Nov 2014.

\bibitem{Dogandzic}
A.~Dogandzic and K.~Qiu, ``{Decentralized Random-Field Estimation for Sensor
  Networks Using Quantized Spatially Correlated Data and Fusion-Center
  Feedback},'' \emph{IEEE Transactions on Signal Processing}, vol.~56, no.~12,
  pp. 6069--6085, 2008.

\bibitem{Epstein}
M.~Epstein, L.~Shi, A.~Tiwari, and R.~M. Murray, ``{Probabilistic performance
  of state estimation across a lossy network},'' \emph{Automatica}, vol.~44,
  no.~12, pp. 3046--3053, Dec. 2008.

\bibitem{Fang}
J.~Fang and H.~Li, ``{Distributed Estimation of Gauss - Markov Random Fields
  With One-Bit Quantized Data},'' \emph{IEEE Signal Processing Letters},
  vol.~17, no.~5, pp. 449--452, 2010.

\bibitem{Dey}
M.~Huang and S.~Dey, ``{Dynamic Quantizer Design for Hidden Markov State
  Estimation Via Multiple Sensors With Fusion Center Feedback},'' \emph{IEEE
  Transactions on Signal Processing}, vol.~54, no.~8, pp. 2887--2896, 2006.

\bibitem{Kreidl}
O.~Kreidl, J.~Tsitsiklis, and S.~Zoumpoulis, ``{On Decentralized Detection With
  Partial Information Sharing Among Sensors},'' \emph{IEEE Transactions on
  Signal Processing}, vol.~59, no.~4, pp. 1759--1765, 2011.

\bibitem{Junlin}
J.~Li and G.~AlRegib, ``{Distributed Estimation in Energy-Constrained Wireless
  Sensor Networks},'' \emph{IEEE Transactions on Signal Processing}, vol.~57,
  no.~10, pp. 3746--3758, 2009.

\bibitem{Chieh}
J.-C. Liu and C.-D. Chung, ``{Distributed Estimation in a Wireless Sensor
  Network Using Hybrid MAC},'' \emph{IEEE Transactions on Vehicular
  Technology}, vol.~60, no.~7, pp. 3424--3435, Sept 2011.

\bibitem{MicheGlobalsip}
N.~Michelusi and U.~Mitra, ``{Distributed estimation in sensor networks with
  quality feedback: A general framework},'' in \emph{IEEE Global Conference on
  Signal and Information Processing (GlobalSIP)}, Dec. 2013, pp. 1057--1060.

\bibitem{MicheAllerton}
------, ``{Fusion center feedback for quasi-decentralized estimation in Sensor
  Networks},'' in \emph{51st Annual Allerton Conference on Communication,
  Control, and Computing (Allerton)}, Oct 2013, pp. 1285--1291.

\bibitem{MicheISIT}
------, ``{A cross-layer framework for joint control and distributed sensing in
  agile wireless networks},'' in \emph{IEEE International Symposium on
  Information Theory (ISIT)}, June 2014, pp. 1747--1751.

\bibitem{MichelusiP2}
------, ``{Cross-layer design of distributed sensing-estimation with quality
  feedback, Part II: Myopic schemes},'' \emph{IEEE Transactions on Signal
  Processing}, 2014.

\bibitem{Msechu}
E.~Msechu and G.~Giannakis, ``{Sensor-Centric Data Reduction for Estimation
  With WSNs via Censoring and Quantization},'' \emph{IEEE Transactions on
  Signal Processing}, vol.~60, no.~1, pp. 400--414, Jan 2012.

\bibitem{Olfati}
R.~Olfati-Saber, ``{Distributed Kalman filtering for sensor networks},'' in
  \emph{46th IEEE Conference on Decision and Control}, 2007, pp. 5492--5498.

\bibitem{Saber}
R.~Olfati-Saber and P.~Jalalkamali, ``{Coupled Distributed Estimation and
  Control for Mobile Sensor Networks},'' \emph{IEEE Transactions on Automatic
  Control}, vol.~57, no.~10, pp. 2609--2614, 2012.

\bibitem{Ray}
P.~Ray and P.~K. Varshney, ``{Distributed Detection in Wireless Sensor Networks
  Using Dynamic Sensor Thresholds},'' \emph{International Journal of
  Distributed Sensor Networks}, vol.~4, no.~1, pp. 5--12, Jan. 2008.

\bibitem{Ribeiro}
A.~Ribeiro and G.~Giannakis, ``{Bandwidth-constrained distributed estimation
  for wireless sensor Networks-part I: Gaussian case},'' \emph{IEEE
  Transactions on Signal Processing}, vol.~54, no.~3, pp. 1131--1143, March
  2006.

\bibitem{Romer}
K.~Romer and F.~Mattern, ``{The design space of wireless sensor networks},''
  \emph{IEEE Wireless Communications}, vol.~11, no.~6, pp. 54--61, 2004.

\bibitem{Saligrama}
V.~Saligrama, M.~Alanyali, and O.~Savas, ``{Distributed Detection in Sensor
  Networks With Packet Losses and Finite Capacity Links},'' \emph{IEEE
  Transactions on Signal Processing}, vol.~54, no.~11, pp. 4118--4132, 2006.

\bibitem{Tandeo}
P.~Tandeo, P.~Ailliot, and E.~Autret, ``\BIBforeignlanguage{English}{{Linear
  Gaussian state-space model with irregular sampling: application to sea
  surface temperature}},'' \emph{\BIBforeignlanguage{English}{{Stochastic
  Environmental Research and Risk Assessment}}}, vol.~25, no.~6, pp. 793--804,
  2011.

\bibitem{Peng}
W.~P. Tay, ``{The Value of Feedback in Decentralized Detection},'' \emph{IEEE
  Transactions on Information Theory}, vol.~58, no.~12, pp. 7226--7239, 2012.

\bibitem{Thatte}
G.~Thatte and U.~Mitra, ``{Sensor Selection and Power Allocation for
  Distributed Estimation in Sensor Networks: Beyond the Star Topology},''
  \emph{IEEE Transactions on Signal Processing}, vol.~56, no.~7, pp.
  2649--2661, 2008.

\bibitem{Tsitsiklis}
J.~N. Tsitsiklis, ``Decentralized detection,'' in \emph{Advances in Statistical
  Signal Processing}.\hskip 1em plus 0.5em minus 0.4em\relax JAI Press, 1993,
  pp. 297--344.

\bibitem{Xiao2}
J.-J. Xiao, S.~Cui, Z.-Q. Luo, and A.~Goldsmith, ``Power scheduling of
  universal decentralized estimation in sensor networks,'' \emph{IEEE
  Transactions on Signal Processing}, vol.~54, no.~2, pp. 413--422, 2006.

\bibitem{Xiao}
J.-J. Xiao, A.~Ribeiro, Z.-Q. Luo, and G.~Giannakis, ``{Distributed
  compression-estimation using wireless sensor networks},'' \emph{IEEE Signal
  Processing Magazine}, vol.~23, no.~4, pp. 27--41, 2006.

\bibitem{Younis}
M.~Younis and K.~Akkaya, ``{Strategies and techniques for node placement in
  wireless sensor networks: A survey},'' \emph{Ad Hoc Networks}, vol.~6, no.~4,
  pp. 621--655, 2008.

\end{thebibliography}

\end{document}